\documentclass[
reprint,
superscriptaddress,
preprintnumbers,
amsmath,
amssymb,
aps,
longbibliography,
pre,
floatfix,
showkeys,
]{revtex4-2}

\usepackage{graphicx}
\usepackage{dcolumn}
\usepackage{bm}
\usepackage{multirow}
\usepackage{textcomp}
\usepackage{gensymb}
\usepackage{lipsum}
\usepackage{longtable}
\usepackage{listings}
\usepackage{booktabs} 
\usepackage{url} 
\usepackage{siunitx} 
\usepackage{graphicx,subfigure,float} 
\usepackage[english]{babel}
\usepackage{gensymb}
\usepackage{fancyhdr}
\usepackage{adjustbox}
\usepackage{mathtools}

\usepackage[T1]{fontenc}
\usepackage[utf8]{inputenc}
\usepackage{amsmath,amssymb,amsfonts,latexsym,cancel}
\usepackage{amsthm} 
\newtheorem{theorem}{Theorem}
\newtheorem{lemma}{Lemma}
\newtheorem{corollary}{Corollary}
\theoremstyle{remark}
\newtheorem{remark}{Remark}
\theoremstyle{plain}
\usepackage[bookmarks]{hyperref}
\usepackage{tabularx}
\usepackage{hhline}
\usepackage{times}
\usepackage{graphicx}
\usepackage{color}
\usepackage{wrapfig}
\usepackage{lmodern}

\usepackage{hyperref}

\usepackage{silence}
\usepackage{color}
\definecolor{crimson}{RGB}{220,20,60}
\definecolor{palevioletred1}{RGB}{225,130,171}
\definecolor{lightskyblue}{RGB}{135,206,250}
\definecolor{gold4}{RGB}{139,117,0}
\definecolor{lightgold1}{RGB}{238,220,130}

\begin{document}

\preprint{APS/123-QED}

\makeatletter
\def\l@subsection#1#2{}
\def\l@subsubsection#1#2{}
\def\l@f@section{}%
\def\toc@@font{\footnotesize \sffamily}%
\makeatother

\title{Environment-Driven Emergence of Higher-Order Collective Behavior}

\author{Felipe S. Abril-Bermúdez}%
\email{felipe.abrilbermudez@abdn.ac.uk}%
\affiliation{School of Natural and Computing Sciences, University of Aberdeen, Aberdeen AB24 3UE, United Kingdom.}%

\author{David N. Fisher}%
\email{david.fisher@abdn.ac.uk}%
\affiliation{School of Biological Sciences, University of Aberdeen, Aberdeen, AB243FX, United Kingdom}

\author{Jean-Baptiste Gramain}%
\email{jbgramain@abdn.ac.uk}%
\affiliation{School of Natural and Computing Sciences, University of Aberdeen, Aberdeen AB24 3UE, United Kingdom.}

\author{Francisco J. Pérez-Reche}%
\email{fperez-reche@abdn.ac.uk}
\affiliation{School of Natural and Computing Sciences, University of Aberdeen, Aberdeen AB24 3UE, United Kingdom.}%

\date{\today}

\begin{abstract}
\noindent
Collective behavior is commonly attributed to direct interactions among system components. We show that higher-order collective structure can instead emerge from shared stochastic environments, even in the absence of interactions. Using the O-information to quantify higher-order dependencies, we prove a no-go theorem for any number of non-interacting variables driven by a single Gaussian environment. According to the theorem, only redundancy can arise when all variables couple to the environment through a common time profile, including constant coupling. Detecting synergy therefore rules out this specific class of environmental mechanisms as a complete explanation, but not the presence of environmental effects in general. In a minimal three-variable model, we show that synergy appears when the coupling profiles differ between variables or when the variables interact directly. With a linear interaction, the model remains exactly solvable and reveals that synergy does not require higher-order interactions. When the variables share a stochastic environment, a simple linear feedback among them is enough. These results identify environmental mediation as a distinct mechanism of higher-order collective organization beyond the conventional interaction-centric paradigm.
\end{abstract}


\maketitle


\section{Introduction}

Collective behavior in complex systems has often been described as emerging from simple local interactions at smaller characteristic scales~\cite{Newman2011, Ladyman2020}. A defining hallmark of complexity, however, is that the behavior of the whole cannot be inferred from its parts, owing to collective dependencies that extend beyond pairwise correlations~\cite{Anderson1972, Goldenfeld1999}. Understanding how such higher-order structure emerges from microscopic interactions remains a central challenge across disciplines~\cite{Matsuda2000, Squartini2018, Prasse2022, Battiston2021, Marinazzo2025,PerezReche2026_HoIReview}.

Early modeling efforts of complex systems focused on phenomena driven by pairwise interactions between elementary units~\cite{Newman2010, Dorogovtsev2022}. However, higher-order interactions, involving dependencies among three or more variables, are now recognized as fundamental to the structure and function of complex systems, from neural activity and biological phenomena to social and financial dynamics~\cite{Matsuda2000, Barrett2015, Rosas2019, Stramaglia2021, Scagliarini2022, Silk2022, Gibbs2022, Gallo2024, Civilini2024, Robiglio2025, Millan2025, Perez-Reche2025}.

Information-theoretic approaches provide a principled framework to connect the interaction mechanisms among elementary units with the collective behaviors that emerge at the system level by quantifying statistical dependencies that go
beyond pairwise correlations~\cite{Marinazzo2025,PerezReche2026_HoIReview}. In particular, the O-information, $\Omega$,
offers a compact measure of the balance between integration and segregation in a system~\cite{Rosas2019}: positive values ($\Omega \! > \!0$) indicate redundancy-dominated regimes, where variables share overlapping information, whereas negative values ($\Omega \! < \! 0$) signal synergy-dominated regimes, where collective effects convey information not present in any part alone---a quantitative expression of Anderson’s “more is different” principle~\cite{Anderson1972}.

The emergence of complex system behavior has typically been described in terms of deterministic interactions among elementary constituents. However, in both natural systems and artificial settings, these interacting units are rarely isolated: neurons, species, or agents often experience common environmental fluctuations that modulate their dynamics and induce correlations or synchronization~\cite{Teramae2004, Sireci2023, Taitelbaum2023, Vasseur2009}. The influence of common environmental inputs on pairwise statistical dependencies is well established in the context of causal inference~\cite{Runge2012, Baldovin2020}, and it has also been addressed in studies of information flow~\cite{Sattari2022} and network reconstruction~\cite{Casadiego2017} in collective dynamics; here we focus instead on how such environments shape higher-order information structure.


We show that shared stochastic environments can themselves generate both redundant and synergistic higher-order behavior, even in the absence of direct interactions between units. We consider a general model of $N$ variables subject to local fluctuations, a shared Gaussian environment, and deterministic interactions, and we quantify higher-order dependencies with the O-information. Depending on the context, the variables may represent species abundances~\cite{Sireci2023,Vasseur2009}, neural membrane potentials~\cite{Moreno-Bote2014}, or coarse-grained magnetization variables (soft spins)~\cite{Onuki2007}.

For non-interacting variables, we prove a no-go theorem that holds for any number of variables and for any Gaussian environment~\cite{Grimmett2001}, whatever its temporal correlations. This class includes white noise, Ornstein--Uhlenbeck processes, and any linear Gaussian process \cite{Sarkka2019}. When all variables couple to a single environment through a common time profile, which includes constant coupling, the environment acts as a hidden common variable and can only produce redundancy. Read in reverse, the theorem provides an exclusion criterion: detecting synergy rules out this broad class of indirect mechanisms as a complete explanation of the observed statistics, whatever the temporal correlations of the Gaussian environment and the size of the system. A common environmental driver may still be present, but other ingredients must then contribute. Settings in which shared environments are expected, such as neural populations driven by common background activity~\cite{Moreno-Bote2014} or species subject to common climatic fluctuations~\cite{Vasseur2009,Sireci2023}, are natural candidates for this test.

To see how synergy arises when the conditions of the theorem are not met, we study a minimal model of three variables, the smallest system for which the O-information is not identically zero. In the non-interacting case, the O-information of this model depends only on the three pairwise correlation coefficients between the variables, which span a three-dimensional correlation space. In this space, a sharp geometric boundary separates redundant and synergistic regimes, providing a direct link between stochastic dynamics~\cite{Gillespie1996}, correlation structure, and emergent higher-order behavior. We find that synergy is more widespread than redundancy in this model, and that whenever the product of the three pairwise correlations is negative, the system is necessarily synergistic.

Synergy appears when the variables couple to the environment with different time profiles, or when they interact directly. For a linear interaction, the model remains exactly solvable, and the transition to synergy has a simple geometric origin: the system moves between two redundant regions of correlation space and must cross the synergistic region on its way. Synergy can therefore emerge from low-order interactions combined with a shared environment, which shows that higher-order behavior need not reflect higher-order mechanisms~\cite{Robiglio2025, Marinazzo2025}.

The paper is organized as follows. Section~\ref{Sec. Model} introduces the general model and the O-information. Section~\ref{Sec. General results} presents general results for non-interacting systems, including the no-go theorem and the exclusion criterion. Section~\ref{Sec. Toy model} analyzes the three-variable model, first without and then with deterministic interactions. Section~\ref{Sec. Conclusions} summarizes our conclusions, and the Appendices contain proofs and derivations.

\section{Model and information measures}
\label{Sec. Model}

\subsection{General model}
\label{Sec. General model}

We consider a system of $N$ variables $\boldsymbol{z}(t)=\left(z_1(t),\ldots,z_N(t)\right)^{T}$ with deterministic initial conditions $z_k(0)$, whose evolution is governed by
\begin{align}
    dz_k(t) = {} & \left[\mu_k + F_k \! \left( \boldsymbol{z} \right)\right] dt + \theta_k\,dW_k(t) \notag \\
    & + f_k(t)\,dA(t), \qquad k=1,\dots,N~.
    \label{Eq. Gen-model}
\end{align}
Each variable is subject to four types of influence. The constant $\mu_k$ is a deterministic drift. The term $\theta_k\,dW_k(t)$, where $\{W_k(t)\}$ are independent standard Wiener processes and $\theta_k \geq 0$ are constants, represents a \emph{local} environment, i.e., idiosyncratic fluctuations acting on $z_k$ alone. The term $f_k(t)\,dA(t)$ represents a \emph{shared} environment: $A(t)$ is a Gaussian process~\cite{Grimmett2001}, independent of $\{W_k\}$ and common to all variables, and the deterministic functions $f_k(t)$ set the strength with which each variable couples to it. Finally, the functions $F_k(\boldsymbol{z})$ describe deterministic interactions between $z_k$ and the other variables.

The statistics of the shared environment enter through its incremental covariance, $K_A(s,r)\,ds\,dr := \mathrm{Cov}\left(dA(s),dA(r)\right)$. The simplest choice is a white-noise environment, $A(t)=W(t)$, for which $K_A(s,r)=\delta(s-r)$; this is the case considered in Sec.~\ref{Sec. Toy model}. Environments with a finite correlation time are described, for instance, by an Ornstein--Uhlenbeck (OU) process: $A(t)=\int_0^t E(s)\,ds$ with $dE=-\gamma E\,dt+\sigma\,dW(t)$, so that $f_k(t)\,dA(t)=f_k(t)\,E(t)\,dt$ and $K_A(s,r)=\mathrm{Cov}\left(E(s),E(r)\right)$, which decays over the time scale $1/\gamma$. More generally, $A(t)$ can be a Gaussian process, including those generated by linear stochastic differential equations driven by Wiener noise. We assume throughout that the stochastic integrals with respect to $A$ exist as Gaussian random variables with finite second moments, and that their covariances are given by the corresponding integrals of the incremental covariance kernel $K_A$. This holds for the white-noise and OU environments considered here.

In the absence of interactions, $F_k(\boldsymbol{z})=0$, Eq.~\eqref{Eq. Gen-model} integrates to
\begin{equation}
    \label{Eq. Gen-solution}
    z_k(t) = z_k(0) + \mu_k t + \theta_k W_k(t) + \int_0^t f_k(s)\,dA(s)~.
\end{equation}
Each $z_k(t)$ is then a linear functional of Gaussian processes, so $\boldsymbol{z}(t)$ is jointly Gaussian~\cite[Sec.~3.3]{Sarkka2019}, $\boldsymbol{z}\sim\mathcal{N}(\boldsymbol{M},\boldsymbol{C})$, with mean $M_k(t)=z_k(0)+\mu_k t+\mathbb{E}\left[\int_0^t f_k(s)\,dA(s)\right]$ and covariance matrix
\begin{equation}
    \label{Eq. Gen-cov-full}
    C_{kl}(t) = \delta_{kl}\,\theta_k^2\,t + \int_0^t\!\int_0^t f_k(s)\,f_l(r)\,K_A(s,r)\,ds\,dr~.
\end{equation}
The first term, due to the local environments, is diagonal; the second term, due to the shared environment, is the only source of statistical dependence between different variables. Interactions that are linear in $\boldsymbol{z}$ preserve Gaussianity, and the covariance matrix can then be obtained exactly (App.~\ref{App. Linear}). Interactions $F_k(\boldsymbol{z})$ that are nonlinear in $\boldsymbol{z}$, such as some of those considered in Sec.~\ref{Sec. Toy model}, make $\boldsymbol{z}$ non-Gaussian; in that case we rely on numerical simulations.

\subsection{O-information}
\label{Sec. O-information}

We quantify higher-order statistical dependencies among the $N$ variables with the O-information~\cite{Rosas2019},
\begin{equation}
    \label{Eq. O-info definition}
    \Omega(\boldsymbol{z}) := \mathrm{TC}(\boldsymbol{z}) - \mathrm{DTC}(\boldsymbol{z})~,
\end{equation}
where
\begin{align}
    \mathrm{TC}(\boldsymbol{z}) &= \sum_{k=1}^{N} H(z_k) - H(\boldsymbol{z})~, \label{Eq. TC}\\
    \mathrm{DTC}(\boldsymbol{z}) &= H(\boldsymbol{z}) - \sum_{k=1}^{N} H(z_k\,|\,\boldsymbol{z}_{-k})~, \label{Eq. DTC}
\end{align}
are the total correlation and the dual total correlation, respectively. Here, $H(\boldsymbol{x})=-\int p(\boldsymbol{x})\log_2 p(\boldsymbol{x})\,d\boldsymbol{x}$ denotes the differential entropy of a random vector $\boldsymbol{x}$ with probability density $p(\boldsymbol{x})$~\cite{Cover2006}, $H(\cdot\,|\,\cdot)$ the conditional differential entropy, and $\boldsymbol{z}_{-k}$ the set of all variables except $z_k$. The total correlation quantifies the overall statistical dependence among the variables, while the dual total correlation quantifies the information shared by at least two of them~\cite{Rosas2019}. Using $H(z_k\,|\,\boldsymbol{z}_{-k}) = H(\boldsymbol{z}) - H(\boldsymbol{z}_{-k})$, Eq.~\eqref{Eq. O-info definition} can be written as
\begin{equation}
    \label{Eq. O-info entropies}
    \Omega(\boldsymbol{z}) = (N-2)\,H(\boldsymbol{z}) + \sum_{k=1}^{N}\left[H(z_k) - H(\boldsymbol{z}_{-k})\right]~.
\end{equation}
As anticipated in the Introduction, $\Omega>0$ indicates redundancy-dominated dependencies and $\Omega<0$ synergy-dominated ones. For $N=3$, $\Omega$ reduces to the co-information, $\Omega = I(z_1;z_2) - I(z_1;z_2\,|\,z_3)$, which is symmetric under permutations of the three variables~\cite{Rosas2019}. In general, $\Omega$ treats all variables on an equal footing and does not require designating a target variable, in contrast to Partial Information Decomposition~\cite{Williams2010}. In the systems considered here, no variable plays a privileged role, even when interactions act on a single variable, which makes $\Omega$ the natural measure.

\section{General results for systems without interactions}
\label{Sec. General results}

In this section, we switch off the deterministic interactions, $F_k(\boldsymbol{z})=0$, so that $\boldsymbol{z}$ is jointly Gaussian with covariance matrix given by Eq.~\eqref{Eq. Gen-cov-full}. Since the entropies of Gaussian variables depend only on their covariances, Eq.~\eqref{Eq. O-info entropies} implies that $\Omega$ is a function of $\boldsymbol{C}$ alone (Lemma~\ref{lem:gaussian} in App.~\ref{App. No-go proof}):
\begin{equation}
    \label{Eq. Gaussian Omega N}
    2\Omega = (N-2)\log_2\det\boldsymbol{C} + \sum_{k=1}^{N}\log_2 C_{kk} - \sum_{k=1}^{N}\log_2\det\boldsymbol{C}_{-k}~,
\end{equation}
where $\boldsymbol{C}_{-k}$ is the covariance matrix of $\boldsymbol{z}_{-k}$, obtained by deleting row and column $k$ of $\boldsymbol{C}$.
\subsection{Common coupling profile and rank-one covariance}
\label{Sec. Common profile}

We say that the environmental couplings share a \emph{common profile} if
\begin{equation}
    \label{Eq. common profile}
    f_k(t) = \varphi_k\, g(t)~, \qquad k=1,\ldots,N~,
\end{equation}
with constants $\varphi_k\in\mathbb{R}$ and a deterministic function $g(t)$ that is the same for all variables. Constant coupling, $f_k(t)=\varphi_k$, corresponds to $g\equiv1$. Under Eq.~\eqref{Eq. common profile}, the shared-environment term in Eq.~\eqref{Eq. Gen-solution} becomes $\varphi_k\tilde{A}(t)$, where
\begin{equation}
    \label{Eq. Atilde}
    \tilde{A}(t) := \int_0^t g(s)\,dA(s)
\end{equation}
is a Gaussian process (for $A=W$, this is an It\^{o} integral, defined for square-integrable $g$; for an OU environment, it is the ordinary integral $\int_0^t g(s)E(s)\,ds$). The covariance matrix~\eqref{Eq. Gen-cov-full} then takes the form
\begin{equation}
    \label{Eq. Gen-cov}
    C_{kl}(t) = \delta_{kl}\,\theta_k^2\,t + \varphi_k\varphi_l\,V_{\mathrm{env}}(t)~, \qquad V_{\mathrm{env}}(t) := \mathrm{Var}\bigl[\tilde{A}(t)\bigr]~.
\end{equation}
The environmental contribution, $V_{\mathrm{env}}\,\boldsymbol{\varphi}\boldsymbol{\varphi}^{T}$ with $\boldsymbol{\varphi}=(\varphi_1,\ldots,\varphi_N)^T$, is a rank-one matrix. Physically, each variable is a rescaled, noisy copy of the same hidden signal $\tilde{A}(t)$: up to its mean, $z_k(t) = \varphi_k\tilde{A}(t) + \theta_k W_k(t)$. In the language of statistics, this is a one-factor model, with the shared environment acting as a hidden common variable. The question is then which kind of higher-order structure such a hidden variable can induce.

\subsection{No-go theorem}
\label{Sec. No-go theorem}

\begin{theorem}[No-go theorem for synergy under a common coupling profile]
\label{thm:nogo}
Consider $N\geq3$ non-interacting variables, $F_k(\boldsymbol{z})=0$, governed by Eq.~\eqref{Eq. Gen-model} with couplings sharing a common profile, Eq.~\eqref{Eq. common profile}. Let $t>0$ be such that $V_{\mathrm{env}}(t)>0$, and assume that the local noise amplitudes satisfy $\theta_k>0$ for all $k$, except possibly for a single variable $m$ with $\theta_m=0$, in which case $\varphi_m\neq0$. Then $\Omega(\boldsymbol{z})\geq0$, with equality if and only if at most two of the $\varphi_k$ are nonzero.
\end{theorem}

When all $\theta_k>0$, the O-information admits the closed form
\begin{equation}
    \label{Eq. Gen-Omega-N}
    \Omega = \frac{1}{2}\log_2\!\left[\frac{(1+S)^{N-2}\prod_{k=1}^{N}\left(1+a_k^2\right)}{\prod_{k=1}^{N}\left(1+S-a_k^2\right)}\right]~,
\end{equation}
where $S=\sum_k a_k^2$ and
\begin{equation}
    \label{Eq. Gen-ak}
    a_k^2(t) = \frac{\varphi_k^2\,V_{\mathrm{env}}(t)}{\theta_k^2\,t}~.
\end{equation}
The proof (App.~\ref{App. No-go proof}) has three steps. The rank-one structure of Eq.~\eqref{Eq. Gen-cov} allows all determinants in Eq.~\eqref{Eq. Gaussian Omega N} to be evaluated with the matrix determinant lemma~\cite{Ding2007}, which yields Eq.~\eqref{Eq. Gen-Omega-N}. Expanding the numerator and denominator in elementary symmetric polynomials~\cite{HornJohnson2013} of the $a_k^2$ then shows that their difference is a sum of non-negative terms, so the argument of the logarithm is never smaller than one. Finally, the case with $\theta_m=0$, which is relevant for the examples of Sec.~\ref{Sec. Toy model}, follows as a limit [Eq.~\eqref{Eq. theta-zero}].

\subsection{Strength of environment-induced redundancy}
\label{Sec. Redundancy strength}

According to Eq.~\eqref{Eq. Gen-Omega-N}, $\Omega$ depends on the model parameters only through the ratios $a_k^2$ between the environmental and local contributions to the variance of $z_k$ accumulated up to time $t$, i.e., through environmental signal-to-noise ratios. The drifts $\mu_k$ and the initial conditions play no role. For a white-noise environment with constant coupling, $a_k=\varphi_k/\theta_k$ and $\Omega$ is constant in time (Corollary~\ref{cor:whitenoise} in App.~\ref{App. Corollaries}, where all corollaries cited below are also proved), as illustrated for the three-variable model in Sec.~\ref{Sec. Toy noninteracting}. For time-dependent profiles $g(t)$, or for environments with finite correlation time such as the OU process, the $a_k$ evolve in time, but $\Omega$ remains non-negative at all times (Corollaries~\ref{cor:whitenoise} and~\ref{cor:OU}).

Two limits clarify how much redundancy a shared environment can induce. At weak coupling ($a_k^2\ll1$ for all $k$),
\begin{equation}
    \label{Eq. weak coupling MT}
    \Omega \simeq \frac{1}{\ln2}\sum_{i<j<k} a_i^2a_j^2a_k^2
\end{equation}
(Corollary~\ref{cor:weak}). Each term equals, to leading order, the O-information of the corresponding triple $\{z_i,z_j,z_k\}$ considered in isolation, so to leading order the redundancy of the whole system is the sum of the redundancies of its triples. This additivity does not hold beyond leading order. For equal ratios, $a_k^2=x$, Eq.~\eqref{Eq. weak coupling MT} becomes $\Omega\simeq\binom{N}{3}x^3/\ln2$ (Corollary~\ref{cor:equal}), which grows with the number $\binom{N}{3}$ of triples, that is, approximately as $N^3$ for large $N$. At strong coupling with equal ratios, $a_k^2=x\to\infty$ at fixed $N$, $\Omega\simeq\frac{N-2}{2}\log_2 x$ (Corollary~\ref{cor:equal}). Redundancy then grows logarithmically with the coupling strength, with a prefactor that increases linearly with system size. In both limits, comparing systems of different size at the same coupling strength shows that the redundancy induced by a single shared environment grows with $N$.

\subsection{Beyond the theorem: routes to synergy and an exclusion criterion}
\label{Sec. Exclusion}

The proof of the no-go theorem uses the dynamics only through the rank-one structure of the covariance matrix, so $\Omega\geq0$ holds for any Gaussian vector whose covariance consists of independent fluctuations for each variable plus a single shared component (Remark~\ref{rem:scope}). Synergy therefore requires that at least one assumption of Theorem~\ref{thm:nogo} fails: (i) the couplings do not share a common profile, (ii) more than one independent shared environment is present, (iii) the shared environment is non-Gaussian, or (iv) the variables interact directly, $F_k(\boldsymbol{z})\neq0$. Routes (i) and (ii) raise the rank of the environmental covariance. For a white-noise environment, common profiles are exactly the couplings that produce a rank-one environmental covariance (Remark~\ref{rem:rank}). Section~\ref{Sec. Toy model} shows that routes (i) and (iv) do produce synergy, through coupling profiles that differ between variables (Sec.~\ref{Sec. Toy noninteracting}) and through deterministic interactions under constant coupling (Sec.~\ref{Sec. Toy interacting}). Violations (ii) and (iii) remove the guarantee of redundancy, but whether they actually generate synergy is beyond the scope of this work.

By contraposition, Theorem~\ref{thm:nogo} provides an \emph{exclusion criterion}. Detected synergy rules out the whole class of mechanisms in which a single Gaussian environment acts on non-interacting variables through a common coupling profile, irrespective of the temporal correlations of the environment, of the profile $g(t)$, and of $N$. More precisely, it rules out such a mechanism as a complete explanation of the joint statistics of the variables. It does not rule out the presence of a common environmental driver, which may act together with other ingredients, such as interactions or additional drivers, that produce the synergy. If, however, independent knowledge of the system establishes that a single Gaussian environment couples to all variables through a common profile, the only assumption of Theorem~\ref{thm:nogo} that can fail is the absence of interactions, so detected synergy reveals direct interactions among the variables. For instance, if synergy were detected in a neural population driven by shared background activity~\cite{Moreno-Bote2014}, or among species subject to common climatic fluctuations~\cite{Vasseur2009,Sireci2023}, at least one of the assumptions of Theorem~\ref{thm:nogo} would have to fail. Heterogeneous coupling profiles and direct interactions are two such violations that we show to produce synergy (Sec.~\ref{Sec. Toy model}); several independent shared drivers and non-Gaussian forcing also lie outside the scope of the theorem, but whether and when they generate synergy remains to be established. The converse does not hold: redundancy is compatible with the theorem but does not confirm its assumptions, since interactions can also produce redundancy (Sec.~\ref{Sec. Toy model}). Results of this kind, which exclude specific indirect mechanisms rather than confirm a direct one, are a realistic target for observational work.

\section{Results for three-variable model}
\label{Sec. Toy model}

\begin{figure}[ht]
    \centering
    \includegraphics[width=5.5cm]{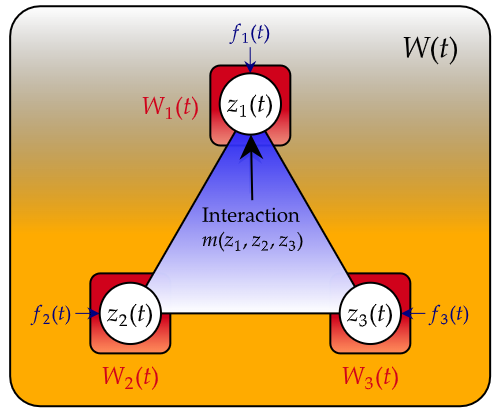}
    \caption{System composed of three elements with state variables $z_{1}(t)$, $z_{2}(t)$, and $z_{3}(t)$. The evolution of all elements is affected by a shared global environment represented by a stochastic process $W(t)$ and by local environment factors $W_{1}(t)$, $W_{2}(t)$, and $W_{3}(t)$. The functions $\left\{f_{k}(t)\right\}_{k=1}^{3}$ quantify the coupling strength between each element and the global environment. The dynamics of the first element include a deterministic interaction term $m \! \left( \boldsymbol{z} \right)$ capturing the interaction with itself and the other two elements.}
    \label{Fig_0.0}
\end{figure}

 We now specialize Eq.~\eqref{Eq. Gen-model} to $N=3$ variables, a white-noise shared environment, $A(t)=W(t)$, and deterministic interactions acting on $z_1$ only, $F_1(\boldsymbol{z})=m(\boldsymbol{z})$ and $F_2=F_3=0$ (see Fig.~\ref{Fig_0.0}):
\begin{align}
    \label{Eq. Toy Model 1}
    dz_{1}(t) \! &= \! \mu_{1}dt + \theta_{1}dW_{1}(t) + f_{1}(t)dW \! (t) + m \! \left( \boldsymbol{z} \right)dt, \\
    \label{Eq. Toy Model 2}
    dz_{2}(t) \! &= \! \mu_{2}dt + \theta_{2}dW_{2}(t) + f_{2}(t)dW \! (t), \\
    \label{Eq. Toy Model 3}
    dz_{3}(t) \! &= \! \mu_{3}dt + \theta_{3}dW_{3}(t) + f_{3}(t)dW \! (t),
\end{align}

This is a minimal setting for the questions addressed here: $N=3$ is the smallest system for which $\Omega$ is not identically zero, restricting interactions to a single variable isolates their role, and the white-noise environment makes the non-interacting case exactly solvable (Sec.~\ref{Sec. General results}).
The function $m \! \left( \boldsymbol{z} \right)$ encodes a deterministic interaction mechanism affecting $z_1(t)$, defined as follows:

\begin{equation}
    \label{Eq. Toy Model 4}
    m \! \left( \boldsymbol{z} \right) \! = \! m_{1}e_{1} \! \left( \boldsymbol{z} \right) \! + \! m_{2}e_{2} \! \left( \boldsymbol{z} \right) \! + \! m_{3}e_{3} \! \left( \boldsymbol{z} \right)~.
\end{equation}
\noindent Here, $e_{1} \! \left( \boldsymbol{z} \right) \! = \! z_{1} \! + \! z_{2} \! + \! z_{3}$, $e_{2} \! \left( \boldsymbol{z} \right) \! = \! z_{1}z_{2} \! + \! z_{1}z_{3} \! + \! z_{2}z_{3}$, and $e_{3} \! \left( \boldsymbol{z} \right) \! = \! z_{1}z_{2}z_{3}$ are the elementary symmetric polynomials. The parameters $m_{1}, \! m_{2}, \! m_{3} \! \in \! \mathbb{R}$ decompose deterministic interactions into single (additive), pairwise, and genuine triplet interactions among the stochastic variables, respectively.

For $m_2\neq0$ or $m_3\neq0$, the interaction term $m(\boldsymbol{z})$ is nonlinear in $\boldsymbol{z}$,

 precluding a closed-form analytical solution.

We considered two complementary approaches. (\textit{i}) Analytical results for the non-interacting case, $m(\boldsymbol{z})=0$, in which $\boldsymbol{z}$ is Gaussian with covariance given by Eq.~\eqref{Eq. Gen-cov-full} (Sec.~\ref{Sec. General model}) and $\Omega$ follows from Eq.~\eqref{Eq. Gaussian Omega N}, and for the linear interaction obtained with $m_2=m_3=0$, for which $\boldsymbol{z}$ remains Gaussian (App.~\ref{App. Linear}). (\textit{ii}) Numerical simulations, which allow individual realizations of the stochastic dynamics to be examined and are required when $m_2\neq0$ or $m_3\neq0$. These are obtained by numerically integrating the stochastic system in Eqs.~\eqref{Eq. Toy Model 1}–~\eqref{Eq. Toy Model 3} using the Euler--Maruyama algorithm~\cite{Kloeden1992}. The O-information in the numerical simulations was estimated using the \textit{HOI} Python toolbox~\cite{Neri2024} with its Gaussian-copula estimator~\cite{Ince2017}. This estimator is exact for Gaussian variables and provides an approximation when nonlinear interactions make the distribution of $\boldsymbol{z}$ non-Gaussian.
\vspace{0.2 cm}

\subsection{Effects of fluctuating environments in non-interacting systems}
\label{Sec. Toy noninteracting}

For $m(\boldsymbol{z})=0$, $\boldsymbol{z}$ is Gaussian (Sec.~\ref{Sec. General model}), and for the white-noise environment considered here the It\^{o} isometry reduces the covariance matrix~\eqref{Eq. Gen-cov-full} to
\begin{align}
    \label{Eq. Fokker-Planck 2 MT}
    C_{kl} (t) \! &= \! \delta_{kl}\theta_{k}^{2} t \! + \! \int_{0}^{t} f_{k}(\tau) f_{l}(\tau) d\tau.
\end{align}

The O-information then follows from Eq.~\eqref{Eq. Gaussian Omega N}, which for three variables can be written in terms of correlation coefficients (App.~\ref{App. Gaussian O-info}):
\begin{equation}
    \label{Eq. Gaussian O-info 5 MT}
    \Omega(\boldsymbol{z})=\frac{1}{2}\log_{2}{ \! \left(\frac{1+2\rho_{12}\rho_{13}\rho_{23}-\rho_{12}^{2}-\rho_{13}^{2}-\rho_{23}^{2}}{\left(1-\rho_{12}^{2}\right) \left(1-\rho_{13}^{2}\right)\left(1-\rho_{23}^{2}\right)} \right)}~,
\end{equation}
\noindent where $\rho_{ij}(t)=C_{ij}/\sqrt{C_{ii}C_{jj}}$ are correlation coefficients.

The condition $\Omega \! = \! 0$ induces a geometric partition of the correlation space $\left[ -1, 1\right]^{3}$ into regions of redundant ($\Omega \! > \! 0$) and synergistic ($\Omega \! < \! 0$) collective behavior. In practice, the zero-O-information manifold is conveniently expressed as the implicit condition $g \! \left[\rho_{12}, \rho_{13}, \rho_{23}\right]=0$, where
\begin{align}
    g \! \left[\rho_{12}, \rho_{13}, \rho_{23}\right] \! &= \! 2\rho_{12}\rho_{13}\rho_{23} \! - \! \rho_{12}^{2}\rho_{13}^{2} \! - \! \rho_{12}^{2}\rho_{23}^{2} \notag\\
    \label{Eq. Fokker-Planck 1}
    & \hspace*{0.35 cm} - \! \rho_{13}^{2}\rho_{23}^{2} \! + \! \rho_{12}^{2}\rho_{13}^{2}\rho_{23}^{2}~.
\end{align}
This manifold is defined under the constraint $g^{\ast} \! = \! 1 \! + \! 2\rho_{12}\rho_{13}\rho_{23} \! - \! \rho_{12}^{2} \! - \! \rho_{13}^{2} \! - \! \rho_{23}^{2} \! > \! 0$; regions of the correlation space where $g^{\ast} \! \leq \! 0$ correspond to parameter combinations for which $\Omega$ is undefined. Indeed, $g^{\ast} > 0$ is precisely the condition ensuring positive definiteness of the covariance matrix $\boldsymbol{C}$ (App.~\ref{App. Gaussian O-info}).

\begin{figure*}[ht]
    \centering
    \includegraphics[width=14cm]{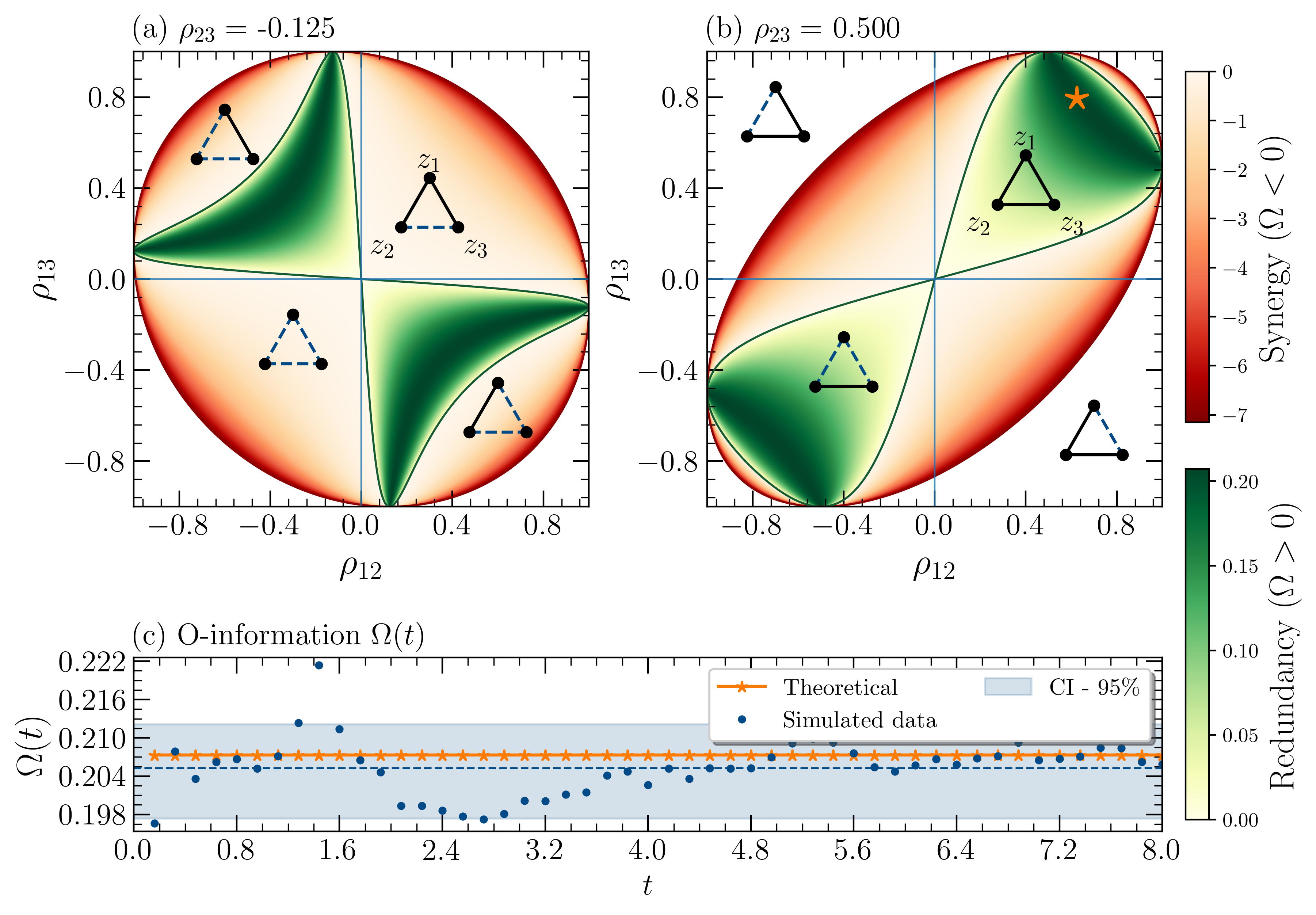}
    \caption{Phase diagram of redundancy (green) and synergy (red) in triplet correlation space. (a,b) Two-dimensional slices of the correlation space $\left( \rho_{12}, \rho_{13}, \rho_{23} \right) \! \in \! \left[ -1, 1\right]^{3}$ at fixed $\rho_{23}$, showing the sign of the O-information: redundant regions ($\Omega \! > \! 0$) and synergistic regions ($\Omega \! < \! 0$). $\Omega$ is not defined in white regions where $g^{\ast} \leq 0$. Triangles show the sign of the pairwise environment-mediated correlations between the dynamical variables $\{z_{1}, z_{2}, z_{3}\}$. Positive correlations, $\rho_{kl}>0$, between $z_{k}$ and $z_{l}$ are marked by continuous black edges. Dotted blue edges mark negative correlations. (c) O-information for a system with constant parameters ($\boldsymbol{\mu} \! = \! \left( 0.5, 0.6, 0.4 \right)^{T}$, $\boldsymbol{\theta} \! = \! \left( 0, 0.25, 0.15 \right)^{T}$ and $\boldsymbol{f}(t) \! = \! \boldsymbol{\varphi} \! = \! (0.2, 0.2, 0.2)^{T}$) yielding the correlations marked by a star in panel (b). The analytical prediction (stars) is compared with numerical simulations (circles). The mean and $95\%$ confidence interval (CI) of the numerical simulations are indicated by the dashed line and shaded region, respectively.}
    \label{Fig_0.1}
\end{figure*}

The partition of the correlation space defined by $g \! = \! 0$ can be visualized through two-dimensional slices at fixed $\rho_{23}$, as shown in Fig.~\ref{Fig_0.1}(a,b). In these sections, the redundant regions form narrow, bow-tie–shaped domains corresponding to specific correlation motifs (indicated by the triangles), while synergistic behavior occupies the complementary, more extended portions of the correlation space where $g^{\ast} \! > \! 0$. The symmetry of these slices under index permutations and even sign flips among the three correlation coefficients directly reflects the invariance of the O-information, as detailed in App.~\ref{App. Gaussian O-info}.

A salient geometric consequence illustrated by the sections in Fig.~\ref{Fig_0.1} is that redundancy is more constrained than synergy. Redundant behavior arises only for specific sign patterns and magnitudes of the correlations: either when all correlations are positive, $\left( \text{sgn}(\rho_{12}), \text{sgn}(\rho_{13}), \text{sgn}(\rho_{23}) \right) \! = \! \left(+, +, + \right)$, or when the system is predominantly anti-correlated, such as $\left( -, -, + \right)$ and its permutations $\mathcal{P} \! \left( -, -, + \right)$.

In contrast, synergy is admissible for all sign combinations and is the \emph{only} possible behavior in fully anti-correlated motifs $\left( -, -, - \right)$ or in partially correlated arrangements of the form $\mathcal{P} \! \left( +, +, - \right)$. According to Harary’s structural balance theory~\cite{Harary1953}, these motifs are unbalanced: they contain cycles with a negative product of edge signs.

This observation agrees with the recent findings of Caprioglio et al.~\cite{Caprioglio2025}, who argued that unbalanced correlation structures are sufficient for synergy dominance in Gaussian systems. Our results further show that synergistic behavior can also arise in balanced correlation motifs with an even number of negative edges, such as $\mathcal{P} \! \left( -, -, + \right)$, although redundancy is also possible in these structures.

Our analytical results further suggest that partially anti-correlated systems (sign motifs $\mathcal{P} \! \left( -, -, + \right)$) may transition between redundancy and synergy as the correlation magnitudes vary, whereas partially correlated systems with $\mathcal{P} \! \left( +, +, - \right)$ are intrinsically synergistic and never exhibit redundancy.


For constant coupling, $f_k(t)=\varphi_k$, Theorem~\ref{thm:nogo} applies with $g\equiv1$ and $V_{\mathrm{env}}=t$. The O-information is therefore constant in time (Corollary~\ref{cor:whitenoise}) and non-negative: redundancy is the only possible outcome when all three variables are coupled to the shared environment, and $\Omega=0$ if any $\varphi_k=0$. The theorem also covers the case $\theta_1=0$ used in Figs.~\ref{Fig_0.1}(c) and~\ref{Fig_0.2}, for which Eq.~\eqref{Eq. theta-zero} gives $\rho_{23}=\rho_{12}\rho_{13}$ and $\Omega=-\frac{1}{2}\log_2\left(1-\rho_{23}^2\right)\geq0$. When more than one $\theta_k$ vanishes, $\boldsymbol{C}$ is singular and $\Omega$ is undefined, although its limit as the local noise vanishes can be computed (App.~\ref{App. No-go proof}).

The correlation-space picture adds a geometric reading of Theorem~\ref{thm:nogo}. For any common coupling profile, Eq.~\eqref{Eq. Gen-cov} gives $\mathrm{sgn}(\rho_{kl})=\mathrm{sgn}(\varphi_k\varphi_l)$ for $k\neq l$, so that $\rho_{12}\rho_{13}\rho_{23}$ has the sign of $\varphi_1^2\varphi_2^2\varphi_3^2>0$. A single shared environment with a common profile can thus only produce balanced motifs, $(+,+,+)$ or $\mathcal{P}(-,-,+)$, and within these it always lies in the redundant region. In particular, the unbalanced motifs, in which synergy is the only possible behavior, are inaccessible.

Fig.~\ref{Fig_0.1}(c) shows the results for the system with constant environmental couplings yielding the correlations marked by the star in Fig.~\ref{Fig_0.1}(b). Redundant behavior is observed with $\Omega \! = \! 0.207$, as given by Eq.~\eqref{Eq. theta-zero} with $\theta_1=0$, $a_2^2=\varphi_2^2/\theta_2^2$, and $a_3^2=\varphi_3^2/\theta_3^2$. Numerical simulations yield a fluctuating O-information $\Omega(t) \! = \! 0.205 [95\%\text{ CI } 0.197-0.212]$ consistent with the analytical prediction (cf. stars and circles in Fig.~\ref{Fig_0.1}(c)).


Next, we consider time-dependent couplings $f_{k}(t) \! = \! \varphi_{k} t^{\alpha_{k}} e^{-\beta_{k} t}$ with $\alpha_{k} \! \geq \! 0$, and $\beta_{k} \! \geq \! 0$, for $k \! \in \! \{1, 2, 3\}$. This functional form of $f_{k}(t)$ encompasses a broad class of time-varying coupling profiles commonly observed in complex systems~\cite{Bhandary2022, Fenn2011}. The couplings share a common profile, Eq.~\eqref{Eq. common profile}, only if $\alpha_k$ and $\beta_k$ are the same for all $k$; otherwise, Theorem~\ref{thm:nogo} does not apply.

\begin{figure}[ht]
    \centering
    \includegraphics[width=7cm]{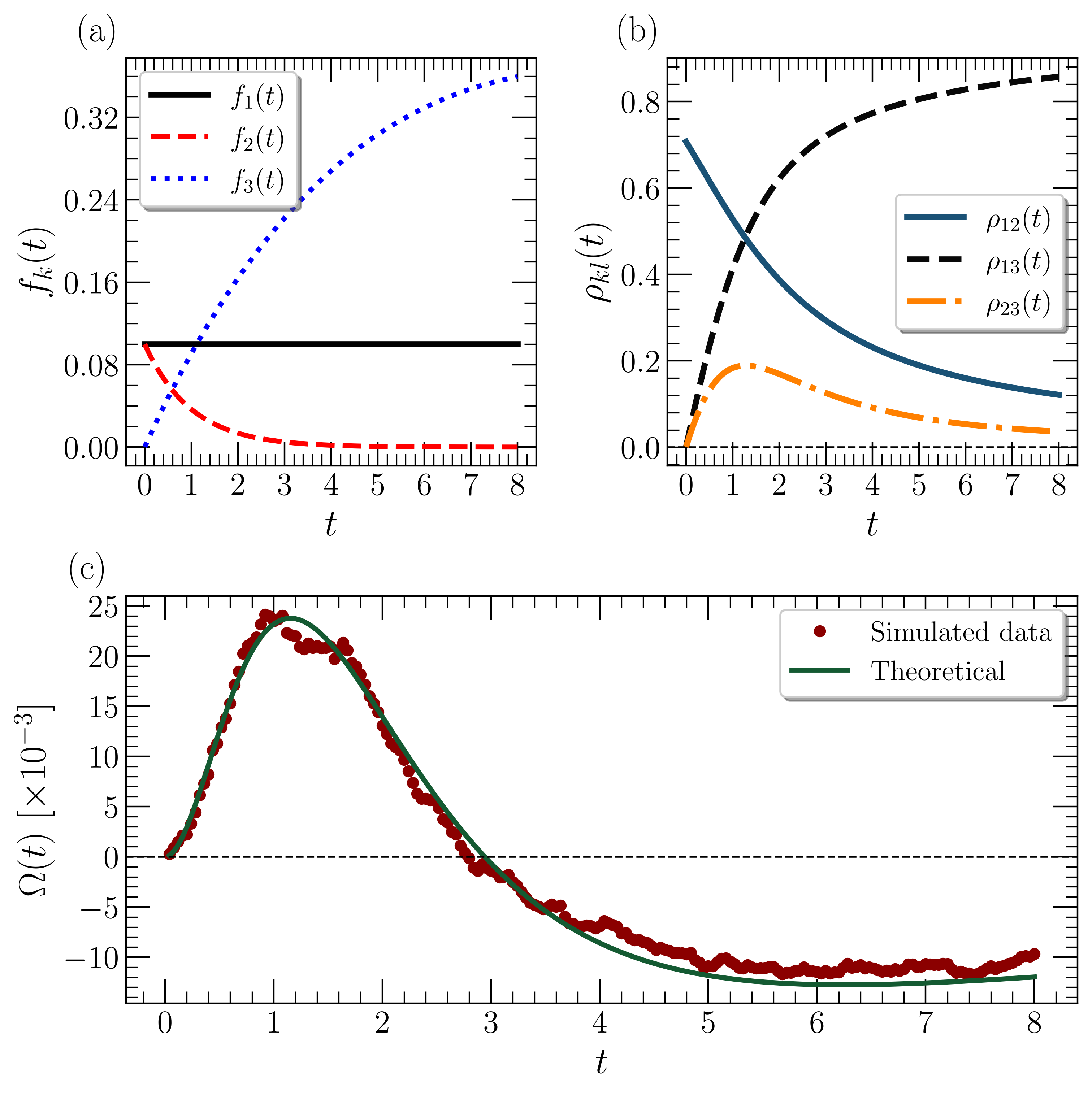}
    \caption{System with time-varying coupling to the shared environment. (a) Temporal variation of the coupling strength to the shared environment for each variable, as indicated in the legend. (b) Time evolution of the pairwise correlation coefficients. (c) Analytical and numerical results for the O-information. Simulations were run with parameters $\boldsymbol{\mu} \! = \! \left( 0.5, 0.6, 0.4 \right)^{T}$, $\boldsymbol{\theta} \! = \! \left( 0, 0.1, 0.1 \right)^{T}$, $\boldsymbol{\varphi} \! = \! \left( 0.1, 0.1, 0.1 \right)^{T}$, $\boldsymbol{\alpha} \! = \! \left( 0, 0, 1 \right)^{T}$ and $\boldsymbol{\beta} \! = \! \left( 0, 1, 0.1 \right)^{T}$, using $2 \! \times \! 10^{4}$ trajectories, and $4 \! \times \! 10^{3}$ time steps.}
    \label{Fig_0.3}
\end{figure}

Our analysis reveals an important finding: \textit{time-dependent coupling to a shared environment with heterogeneous profiles can induce redundancy-to-synergy transitions.} Fig.~\ref{Fig_0.3} illustrates this behavior for a system with varying couplings $f_{2}(t)$ and $f_{3}(t)$ (panel (a)), which drive a dynamical evolution within correlation space (panel (b)). $\Omega(t)$ transitions from positive (redundant) to negative (synergistic) values over time, as shown by both analytical predictions, obtained by inserting the covariance~\eqref{Eq. Fokker-Planck 2 MT} into Eq.~\eqref{Eq. Gaussian O-info 5 MT}, and numerical simulations. Here the three profiles differ, which corresponds to route (i) in Sec.~\ref{Sec. Exclusion}. The excellent agreement between theory and simulations is further documented in the Supplemental Material~\cite{SM}, where we compare the time evolution of the mean vector $\boldsymbol{M}$ and covariance matrix $\boldsymbol{C}$ obtained from analytical and numerical results.

\subsection{Combined effect of environmental coupling and deterministic interactions}
\label{Sec. Toy interacting}

\begin{figure}[ht]
    \centering
    \includegraphics[width=7cm]{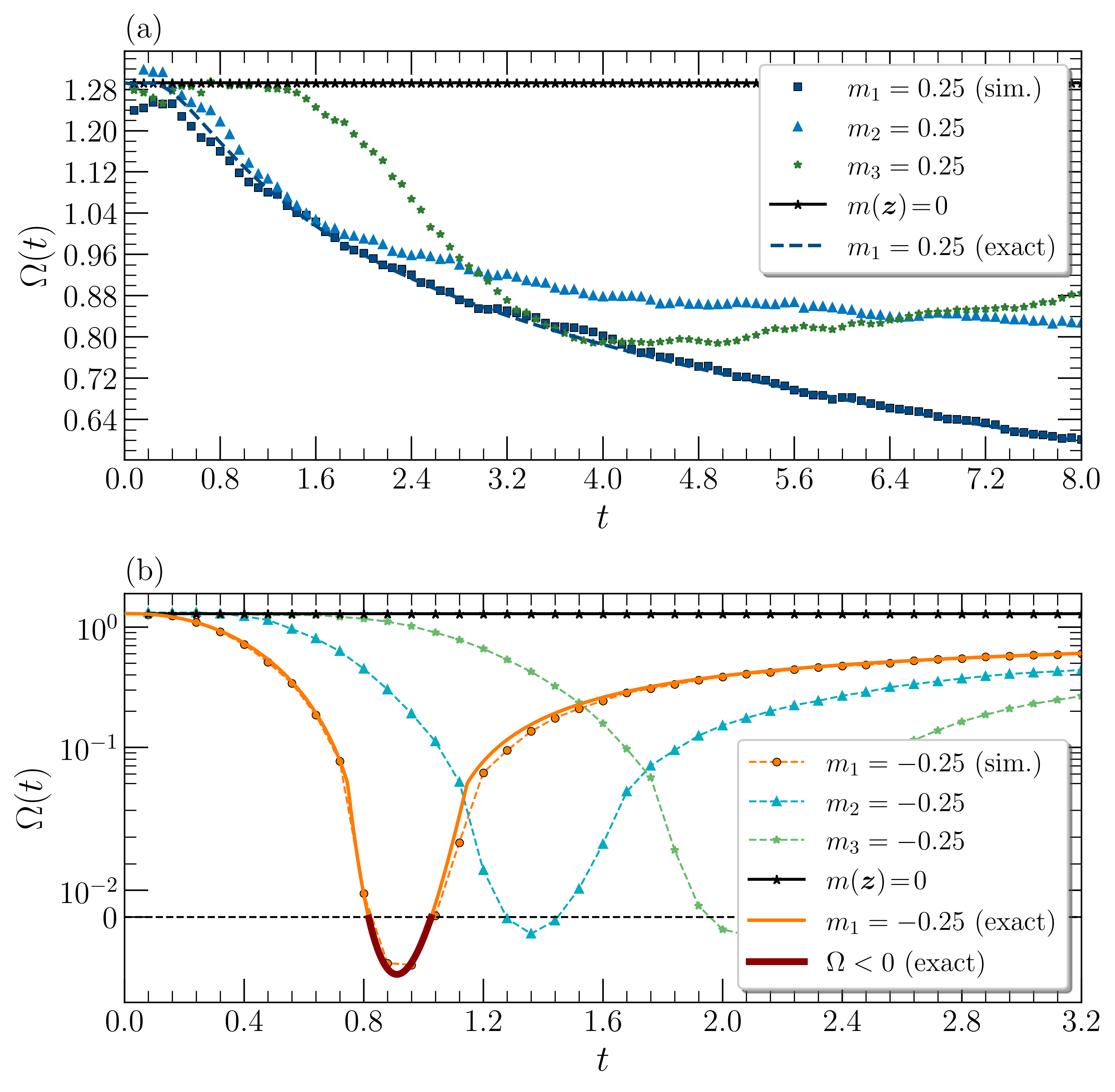}
    \caption{Effect of deterministic interactions on the behavior of systems with \emph{constant} coupling to a shared environment. Time evolution of the O-information for several values of the deterministic couplings $m_{1}$, $m_{2}$, and $m_{3}$, indicated by the symbols in the legends. Panels (a) and (b) correspond to positive and negative values of the deterministic couplings, respectively. The solid black lines show a baseline for $\Omega(t)$ defined by the corresponding systems without deterministic interactions (i.e., with $m \! \left( \boldsymbol{z} \right) \! = \! 0$). For the linear interactions, the exact results (App.~\ref{App. Linear}) are shown by the dashed blue line ($m_1=0.25$) and the solid orange line ($m_1=-0.25$), whose thick dark-red segment marks the synergistic interval; the corresponding simulations are shown by blue squares and orange circles, as in Fig.~\ref{Fig_traj}. Simulations were run with parameters $\boldsymbol{\mu} \! = \! \left( 0.5, 0.6, 0.4 \right)^{T}$, $\boldsymbol{\theta} \! = \! \left( 0, 0.25, 0.15 \right)^{T}$ and $\boldsymbol{f}(t) \! = \! \boldsymbol{\varphi} \! = \! \left( 0.2, 0.6, 1 \right)^{T}$, using $2\times10^{4}$ trajectories, and $4\times10^{3}$ time steps.}
    \label{Fig_0.2}
\end{figure}

\begin{figure*}[ht]
    \centering
    \includegraphics[width=14 cm]{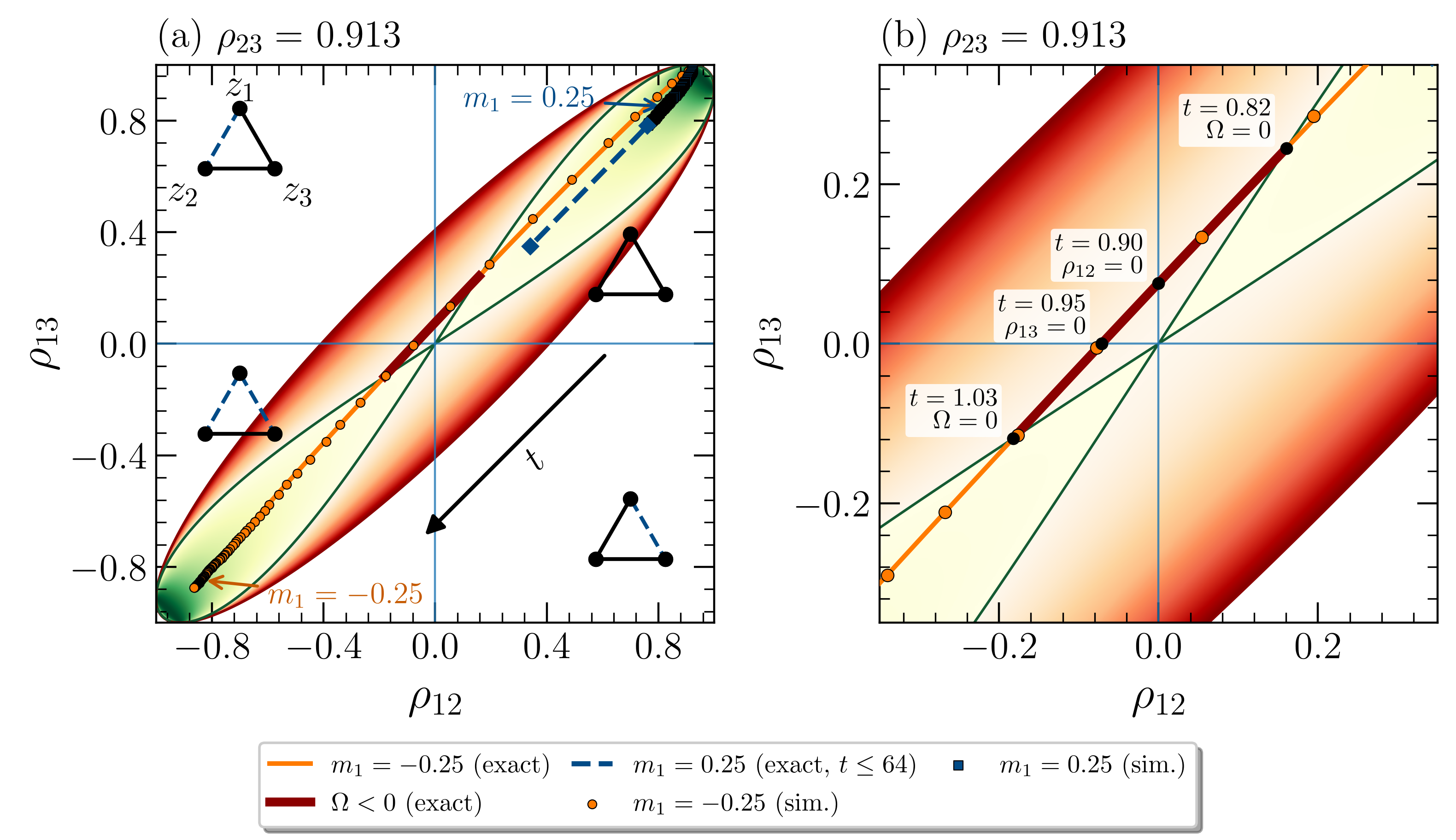}
    \caption{Trajectory in correlation space for the linear interaction $m_1=-0.25$ (with $m_2=m_3=0$). (a) Slice $\rho_{23}\simeq0.913$ of the correlation space, which remains fixed because the interaction does not act on $z_2$ and $z_3$; background colors, contours, and sign triangles as in Fig.~\ref{Fig_0.1}. The orange line is the exact trajectory $(\rho_{12}(t),\rho_{13}(t))$, and its thick dark-red segment marks the synergistic interval $0.82\lesssim t\lesssim1.03$. The dashed blue line is the exact trajectory for $m_1=0.25$, drawn up to $t=64$; the two diamonds mark $t=8$ (end of the simulations, near the starting point) and $t=64$ (end of the line); it approaches the origin slowly without leaving the upper redundant region. Markers show the simulated correlations up to $t=8$ (orange circles: $m_1=-0.25$; blue squares: $m_1=0.25$). The black arrow indicates the direction of increasing time along both trajectories. (b) Zoom on the region near the origin, indicating the times at which $\Omega$, $\rho_{12}$, and $\rho_{13}$ change sign. Other parameters as in Fig.~\ref{Fig_0.2}.}
    \label{Fig_traj}
\end{figure*}

We now switch on the deterministic interactions, which corresponds to route (iv) in Sec.~\ref{Sec. Exclusion}. Our numerical simulations reveal a rich phenomenology arising from the interplay between environmental coupling and deterministic interactions, quantified by $f_{k}(t)$ and $m \! \left( \boldsymbol{z} \right)$, respectively. Fig.~\ref{Fig_0.2} illustrates the time evolution of $\Omega(t)$ for a system with constant coupling to the shared environment, under different choices of the deterministic interaction parameters $m \! \left( \boldsymbol{z} \right)$. Without interactions, Theorem~\ref{thm:nogo} restricts this system to redundancy, with the constant value $\Omega_0\simeq1.29$ given by Eq.~\eqref{Eq. theta-zero} (solid horizontal lines in Fig.~\ref{Fig_0.2}).
We first consider the linear interaction ($m_2=m_3=0$), for which the model remains exactly solvable, and then the nonlinear interactions ($m_2\neq0$ or $m_3\neq0$), which we study numerically.

\emph{Linear interactions.} For $m_2=m_3=0$, the interaction $m(\boldsymbol{z})=m_1(z_1+z_2+z_3)$ is linear, and Eqs.~\eqref{Eq. Toy Model 1}--\eqref{Eq. Toy Model 3} form a linear stochastic differential equation with additive noise. The variables then remain Gaussian, and $\Omega(t)$ follows exactly from the covariance matrix derived in App.~\ref{App. Linear}. The expression for the covariance matrix given in App.~\ref{App. Linear general} is not restricted to three variables and holds for linear interactions among any number of variables. We note that, although we are dealing with three variables, the expression for the covariance matrix given in App.~\ref{App. Linear general} holds for linear interactions among any number of variables. The interaction breaks the structure required by Theorem~\ref{thm:nogo}: through the feedback, $z_1$ integrates the local noises of $z_2$ and $z_3$, and the shared signal enters $z_1$ with a time-dependent weight, so that $\boldsymbol{C}$ no longer has the form of independent fluctuations plus a single shared component required by Theorem~\ref{thm:nogo}. Figure~\ref{Fig_0.2} compares the exact results (solid colored lines) with the simulations. For $m_1=0.25$, $\Omega(t)$ remains positive and decreases to $\Omega\simeq0.60$ at $t=8$; at longer times, the correlations $\rho_{12}$ and $\rho_{13}$ decay slowly while the system stays redundant (Fig.~\ref{Fig_traj}). For $m_1=-0.25$, $\Omega(t)$ is negative for $0.82\lesssim t\lesssim1.03$, with a minimum $\Omega\simeq-0.021$ at $t\simeq0.91$, in agreement with the simulations, which cross zero at $t\simeq0.84$. This is an exactly solvable example in which a shared environment and low-order deterministic interactions jointly produce synergy.

The correlation-space picture of Sec.~\ref{Sec. Toy noninteracting} explains this transition, which can be followed in Fig.~\ref{Fig_traj}. Since the interaction does not act on $z_2$ and $z_3$, $\rho_{23}\simeq0.913$ remains constant, and the trajectory $(\rho_{12}(t),\rho_{13}(t))$ lies in a single slice of the correlation space [Fig.~\ref{Fig_traj}(a)]. For $m_1<0$, the feedback drives $z_1$ towards $-(z_2+z_3)$, gradually eroding its positive correlations with $z_2$ and $z_3$, so that the trajectory moves along the diagonal from the upper-right corner towards the origin. Synergy appears at $t\simeq0.82$, while all three correlations are still positive but $\rho_{12}$ and $\rho_{13}$ are small: $z_1$ is then nearly independent of $z_2$ and of $z_3$ individually, but not of the pair. Between $t\simeq0.90$ and $t\simeq0.95$, $\rho_{12}$ has become negative while $\rho_{13}$ has not, so the system crosses the unbalanced motif $(-,+,+)$, where synergy is the only possible behavior [Fig.~\ref{Fig_traj}(b)]. Once $\rho_{13}$ also becomes negative, the system enters the balanced motif $\mathcal{P}(-,-,+)$, and redundancy returns as the magnitudes of the correlations grow.
In this slice, the two redundant regions touch only at the origin, $\rho_{12}=\rho_{13}=0$. The feedback carries the system from the upper redundant region, where it starts, to the lower one, and any path between them that avoids this single point must cross the synergistic region. The synergistic episode is therefore a geometric consequence of the path rather than a fine-tuned feature, and it is brief because the trajectory passes close to the origin.

\emph{Nonlinear interactions.} The terms $m_2e_2(\boldsymbol{z})$ and $m_3e_3(\boldsymbol{z})$ grow with the magnitudes of $z_2$ and $z_3$, which drift and diffuse. The results for these interactions therefore refer to the finite time window shown in Fig.~\ref{Fig_0.2}, and the effects described below are transient.

Nonzero deterministic interactions ($m \! \left( \boldsymbol{z} \right) \! \neq \! 0$) enable redundancy--synergy transitions even when the environmental couplings favor redundancy. This is illustrated in Figure~\ref{Fig_0.2}(b) for negative deterministic interactions; for $m_2=-0.25$ and $m_3=-0.25$, the transitions occur later than for $m_1=-0.25$, at $t\simeq1.25$ and $t\simeq1.99$, respectively. Conversely, the positive interactions in Fig.~\ref{Fig_0.2}(a) leave the system redundant, so that observed redundancy does not by itself confirm the assumptions of Theorem~\ref{thm:nogo} (Sec.~\ref{Sec. Exclusion}).

Remarkably, synergy can emerge not only from higher-order interactions ($m_3 \! < \! 0$) but also from low-order deterministic terms with negative $m_1$ or $m_2$. The effect is even stronger and occurs earlier in these low-order cases. For the linear interaction, the exact solution shows how: no explicit higher-order term is needed, because the interplay between the feedback and the shared environment reshapes the correlation structure. This illustrates that higher-order behavior need not reflect higher-order mechanisms~\cite{Robiglio2025, Marinazzo2025}: in the presence of a shared environment, synergy can arise from deterministic interactions of low order.

\section{Conclusions}
\label{Sec. Conclusions}

We have shown that shared stochastic environments can generate higher-order collective behavior, both redundant and synergistic, even in the absence of direct interactions among the units of a system. For non-interacting variables driven by a single Gaussian environment, we proved a no-go theorem that holds for any number of variables $N\geq3$ and whatever the temporal correlations of this Gaussian environment. When all variables couple to the environment through a common time profile, the environmental contribution to the covariance matrix has rank one, and the O-information can only be non-negative. In this regime, the shared environment acts as a hidden common variable, of which every unit carries a noisy copy. The closed-form expression for $\Omega$ shows that the resulting redundancy is controlled by the ratios between environmental and local fluctuations, is approximately a sum over triples of variables at weak coupling, and grows with system size.

Synergy requires breaking this structure. In the three-variable model, we identified two routes in which this can happen. First, coupling profiles that differ between variables induce redundancy-to-synergy transitions even without interactions. Second, deterministic interactions also produce synergy, as shown for a linear feedback, for which the model remains exactly solvable. In that case, synergy appears because the system moves between two redundant regions of correlation space that touch only at a single point, so that its path must cross the synergistic region. Synergy can thus arise from low-order interactions combined with a shared environment, which illustrates that higher-order behavior need not reflect higher-order mechanisms~\cite{Robiglio2025, Marinazzo2025}. The signs of the three pairwise correlations offer a simple way to see this. A single shared environment with a common profile can only produce balanced sign patterns, in which the product $\rho_{12}\rho_{13}\rho_{23}$ is positive, such as all three correlations being positive. Synergy, by contrast, occupies larger regions of correlation space, and it is the only possible behavior for unbalanced patterns, in which this product is negative.

The no-go theorem also has a direct observational reading. Detected synergy rules out the whole class of mechanisms in which a single Gaussian environment acts on non-interacting units through a common coupling profile, irrespective of the temporal correlations of the environment and of system size. This exclusion concerns such a mechanism as a complete explanation of the observed statistics. A common environmental driver may still be present, acting together with other ingredients that produce the synergy. In particular, if the shared environment is known to be a single Gaussian driver acting on all units with the same time profile, observed synergy implies that the units interact directly. Shared background activity in neural populations~\cite{Moreno-Bote2014} and common climatic fluctuations acting on species~\cite{Vasseur2009,Sireci2023} are natural settings in which to apply this test: if synergy were detected in such a system, at least one assumption of the theorem would have to be violated. Such tests are already within reach. For example, the O-information of spiking activity recorded from macaque neurons during a decision task was found to be dominated by redundancy, with synergy appearing only in time-lagged dependencies~\cite{Stramaglia2021}. Our theorem shows that this redundancy is compatible with a common environmental drive, but it does not prove that such a drive is present. Indeed, redundancy can also arise when units interact directly [Fig.~\ref{Fig_0.2}(a)], or when they couple to a shared environment with different time profiles (early times in Fig.~\ref{Fig_0.3}). Conversely, if equal-time synergy were found in other neural recordings, it would rule out a single Gaussian drive acting on non-interacting neurons through a common coupling profile as a complete explanation of the data. We have shown that two such violations, heterogeneous coupling profiles and direct interactions, do produce synergy. Several independent environmental drivers and non-Gaussian forcing are further violations; they remove the guarantee of redundancy, but whether and under which conditions they generate synergy remains open. Observing redundancy, on the other hand, does not show that the system is non-interacting and driven by a common environment, since interactions can also produce redundancy. Results of this kind, which exclude specific indirect mechanisms rather than confirm a direct one, are a realistic target for observational work, and they complement approaches that address the effect of common inputs at the level of pairwise dependencies and information flow~\cite{Runge2012, Sattari2022, Casadiego2017}.

Several extensions remain open. Two of them concern the violations of the theorem's assumptions that we have not explored. Our analytical results assume Gaussian shared environments, whereas environmental fluctuations in many systems are not Gaussian, for instance Poisson-like inputs in neural circuits or extreme climatic events in ecological systems~\cite{Taitelbaum2023}. Likewise, the no-go theorem covers a single shared environment and does not, without further argument, extend to several independent ones. Determining whether non-Gaussian or multiple environments can generate synergy, and under which conditions, would extend this picture of environment-driven higher-order behavior beyond the cases treated here. In this study, interacting systems were treated exactly only for a linear feedback among three variables, so nonlinear interactions and larger interacting systems call for further work. Finally, we characterized higher-order structure with the O-information; assessing the robustness of the redundancy--synergy picture with respect to alternative measures of higher-order dependence is a natural next step.

More broadly, the coexistence of environmental and interaction-driven mechanisms that produce similar collective behavior implies that observations of the behaviour alone may be insufficient to identify its origin. Methods capable of disentangling these contributions will be important for understanding, and eventually controlling, collective behavior in complex systems.

\section{Acknowledgments}
This research was funded by the BBSRC grant no. BB/Y513027/1.

\section{Data availability} The data that support the findings of this article are publicly available~\cite{GithubOinfo2025}.

\appendix

\section{Mathematical results for non-interacting systems}
\label{App. No-go proof}

This appendix collects the mathematical results for non-interacting systems presented in Sec.~\ref{Sec. General results}. The argument proceeds in three steps. First, Lemma~\ref{lem:gaussian} (App.~\ref{App. Gaussian lemma}) expresses the O-information of any set of jointly Gaussian variables in terms of determinants of their covariance matrix. Second, the proof of Theorem~\ref{thm:nogo} (App.~\ref{App. Theorem proof}) applies this lemma to the covariance matrix obtained when all variables share a common coupling profile. Its environmental part then has rank one, and the resulting expression for the O-information can be shown never to be negative. Finally, the remarks (App.~\ref{App. Remarks}) discuss limiting cases and the scope of the theorem, and the corollaries (App.~\ref{App. Corollaries}) apply it to specific environments and coupling regimes.

\subsection{O-information of Gaussian variables}
\label{App. Gaussian lemma}

\begin{lemma}[Gaussian O-information]
\label{lem:gaussian}
Let $\boldsymbol{z}=(z_1,\ldots,z_N)^T$, $N\geq2$, be jointly Gaussian with positive-definite covariance matrix $\boldsymbol{C}$. Then $\Omega(\boldsymbol{z})$ is given by Eq.~\eqref{Eq. Gaussian Omega N}. In particular, $\Omega$ does not depend on the mean of $\boldsymbol{z}$.
\end{lemma}

\begin{proof}
The differential entropy of an $n$-dimensional random vector $\boldsymbol{x}$ with probability density $p(\boldsymbol{x})$ is $H(\boldsymbol{x})=-\int p(\boldsymbol{x})\log_2 p(\boldsymbol{x})\,d\boldsymbol{x}=-\mathbb{E}[\log_2 p(\boldsymbol{x})]$~\cite{Cover2006}. For a Gaussian vector with mean $\boldsymbol{M}$ and positive-definite covariance $\boldsymbol{\Sigma}$,
\begin{align*}
  -\log_2 p(\boldsymbol{x}) = {} & \tfrac{1}{2}\log_2\bigl[(2\pi)^n\det\boldsymbol{\Sigma}\bigr] \\
  & + \frac{(\boldsymbol{x}-\boldsymbol{M})^T\boldsymbol{\Sigma}^{-1}(\boldsymbol{x}-\boldsymbol{M})}{2\ln2}~.
\end{align*}
The expectation of the quadratic form is $\mathrm{tr}\bigl(\boldsymbol{\Sigma}^{-1}\,\mathbb{E}[(\boldsymbol{x}-\boldsymbol{M})(\boldsymbol{x}-\boldsymbol{M})^T]\bigr)=\mathrm{tr}(\mathbb{I}_n)=n$, and $n/(2\ln2)=\tfrac{1}{2}\log_2 e^n$. Hence
\begin{equation}
  \label{Eq. Gaussian entropy}
  H(\boldsymbol{x}) = \tfrac{1}{2}\log_2\bigl[(2\pi e)^n\det\boldsymbol{\Sigma}\bigr]~,
\end{equation}
which depends on $\boldsymbol{\Sigma}$ but not on $\boldsymbol{M}$~\cite{Cover2006}. Every marginal of a Gaussian vector is Gaussian, and the covariance of $\boldsymbol{z}_{-k}$ is the principal submatrix $\boldsymbol{C}_{-k}$, which is positive definite. Applying Eq.~\eqref{Eq. Gaussian entropy} to $\boldsymbol{z}$, $z_k$, and $\boldsymbol{z}_{-k}$,
\begin{align*}
  H(\boldsymbol{z}) &= \tfrac{1}{2}\log_2\bigl[(2\pi e)^N\det\boldsymbol{C}\bigr],\\
  H(z_k) &= \tfrac{1}{2}\log_2\bigl[2\pi e\,C_{kk}\bigr],\\
  H(\boldsymbol{z}_{-k}) &= \tfrac{1}{2}\log_2\bigl[(2\pi e)^{N-1}\det\boldsymbol{C}_{-k}\bigr].
\end{align*}
Substituting into Eq.~\eqref{Eq. O-info entropies}, the coefficient of $\log_2(2\pi e)$ in $2\Omega$ is $N(N-2)+N-N(N-1)=0$, which leads to Eq.~\eqref{Eq. Gaussian Omega N}.
\end{proof}

\subsection{Proof of Theorem~\ref{thm:nogo}}
\label{App. Theorem proof}

\textit{Gaussianity and covariance.} With $F_k=0$ and Eq.~\eqref{Eq. common profile}, Eq.~\eqref{Eq. Gen-model} integrates to $z_k(t)=z_k(0)+\mu_k t+\theta_k W_k(t)+\varphi_k\tilde{A}(t)$. The process $\tilde{A}$ is a linear functional of the Gaussian process $A$, hence Gaussian, and it is independent of $\{W_k\}$. The vector $(W_1,\ldots,W_N,\tilde{A})$ is therefore jointly Gaussian, and so is $\boldsymbol{z}(t)$, with mean $M_k(t)=z_k(0)+\mu_k t+\varphi_k\,\mathbb{E}[\tilde{A}(t)]$ and covariance given by Eq.~\eqref{Eq. Gen-cov}. In matrix form, $\boldsymbol{C}=\boldsymbol{D}+V_{\mathrm{env}}\,\boldsymbol{\varphi}\boldsymbol{\varphi}^T$ with $\boldsymbol{D}=t\,\mathrm{diag}(\theta_1^2,\ldots,\theta_N^2)$.

\textit{Reduction (all $\theta_k>0$).} By the matrix determinant lemma~\cite{Ding2007}, $\det\boldsymbol{C}=\det\boldsymbol{D}\,(1+V_{\mathrm{env}}\boldsymbol{\varphi}^T\boldsymbol{D}^{-1}\boldsymbol{\varphi})=\bigl(\prod_k\theta_k^2t\bigr)(1+S)$. Each submatrix $\boldsymbol{C}_{-k}$ has the same structure as $\boldsymbol{C}$ with variable $k$ removed, so $\det\boldsymbol{C}_{-k}=\bigl(\prod_{j\neq k}\theta_j^2t\bigr)(1+S-a_k^2)$. Also, $C_{kk}=\theta_k^2t\,(1+a_k^2)$. Substituting into Eq.~\eqref{Eq. Gaussian Omega N}, each $\log_2(\theta_j^2t)$ appears with coefficient $(N-2)+1-(N-1)=0$, which leads to Eq.~\eqref{Eq. Gen-Omega-N}.

\textit{Positivity.} Write the argument of the logarithm in Eq.~\eqref{Eq. Gen-Omega-N} as $b/c$, with
\begin{equation*}
  b=(1+S)^{N-2}\prod_k(1+a_k^2)~,\qquad c=\prod_k(1+S-a_k^2)~.
\end{equation*}
Each factor of $c$ equals $1+\sum_{j\neq k}a_j^2\geq1$, so $c\geq1$, and $\Omega=\frac{1}{2}\log_2\left[1+(b-c)/c\right]$. It is therefore enough to show that $b-c\geq0$. Let $e_j$ denote the $j$-th elementary symmetric polynomial in $a_1^2,\ldots,a_N^2$~\cite[Definition~1.2.14]{HornJohnson2013}, that is, the sum of all products of $j$ distinct factors $a_k^2$,
\begin{equation*}
  e_j=\sum_{1\leq k_1<\cdots<k_j\leq N}a_{k_1}^2\cdots a_{k_j}^2~,
\end{equation*}
with the convention $e_0=1$. In particular, $e_1=\sum_k a_k^2=S$. Since all $a_k^2$ are non-negative, every $e_j$ is non-negative as well. Expanding,
\begin{align*}
  \prod_k(1+a_k^2) &= \sum_{j=0}^N e_j~,\\
  \prod_k\bigl[(1+S)-a_k^2\bigr] &= \sum_{j=0}^N(-1)^j(1+S)^{N-j}e_j~,
\end{align*}
hence
\begin{equation}
  b-c=\sum_{j=0}^N e_j\,(1+S)^{N-j}\bigl[(1+S)^{j-2}-(-1)^j\bigr].
  \label{Eq. diff-expanded}
\end{equation}
The $j=2$ term vanishes identically, and the $j=0$ and $j=1$ terms sum to $(1+S)^{N-2}\bigl[1-(1+S)^2+S(2+S)\bigr]=0$. For $j\geq3$, the bracket equals $(1+S)^{j-2}+1\geq2$ for odd $j$ and $(1+S)^{j-2}-1\geq0$ for even $j$. Therefore, only the terms with $j\geq3$ remain in Eq.~\eqref{Eq. diff-expanded}, and all of them are non-negative. Hence $b-c\geq0$, so that $b\geq c$ and $\Omega\geq0$.

\textit{Equality.} All remaining terms in Eq.~\eqref{Eq. diff-expanded} are non-negative and the $j=3$ term equals $e_3(1+S)^{N-3}(2+S)$, so the sum vanishes if and only if $e_3=0$. Since $a_k^2\geq0$, this happens exactly when at most two $a_k^2$ are nonzero, in which case $e_j=0$ for all $j\geq3$. Because $V_{\mathrm{env}}(t)>0$, $a_k=0$ if and only if $\varphi_k=0$.

\textit{Vanishing local noise for one variable.} Suppose $\theta_m=0$ with $\varphi_m\neq0$. For any $\boldsymbol{x}\neq0$, $\boldsymbol{x}^T\boldsymbol{C}\boldsymbol{x}=t\sum_{k\neq m}\theta_k^2x_k^2+V_{\mathrm{env}}(\boldsymbol{\varphi}\cdot\boldsymbol{x})^2>0$, so $\boldsymbol{C}$ is positive definite. By Lemma~\ref{lem:gaussian}, $\Omega$ is a continuous function of $\boldsymbol{C}$ on positive-definite matrices, so it equals the limit $\theta_m\to0$ ($a_m^2\to\infty$) of Eq.~\eqref{Eq. Gen-Omega-N}. The $N-1$ powers of $a_m^2$ that diverge in the numerator and in the denominator cancel, giving
\begin{equation}
  \label{Eq. theta-zero}
  \Omega=\frac12\log_2\!\left[\frac{\prod_{k\neq m}(1+a_k^2)}{1+\sum_{k\neq m}a_k^2}\right]\geq0,
\end{equation}
since $\prod_k(1+x_k)\geq1+\sum_kx_k$ for $x_k\geq0$, with equality if and only if at most one $x_k$ is nonzero. As $\varphi_m\neq0$, equality again means that at most two $\varphi_k$ are nonzero. \qed

\subsection{Remarks}
\label{App. Remarks}

\begin{remark}
For $N=3$, Eq.~\eqref{Eq. diff-expanded} reduces to $b-c=a_1^2a_2^2a_3^2(2+S)\geq0$, and Eq.~\eqref{Eq. theta-zero} with $m=1$ gives $\Omega=-\frac12\log_2(1-\rho_{23}^2)$, with $\rho_{23}=\rho_{12}\rho_{13}$. Let $q$ be the number of variables with $\theta_k=0$ and $\varphi_k\neq0$. For $q\geq2$, these variables are perfectly correlated and $\boldsymbol{C}$ is singular, so the differential-entropy definition of $\Omega$ does not apply directly. One can still take the limit in which the corresponding $\theta_k$ tend to zero, i.e., $a_k^2\to\infty$, in Eq.~\eqref{Eq. Gen-Omega-N}. Counting the diverging factors in the numerator and the denominator shows that the argument of the logarithm grows as $(a^2)^{q-2}$, where $a^2$ denotes the common scale of the diverging $a_k^2$. For $q=2$, the limit is finite and independent of how the two $a_k^2$ diverge,
\begin{equation}
  \label{Eq. theta-zero-two}
  \Omega\to\frac12\log_2\prod_{k\notin\mathcal{Z}}\left(1+a_k^2\right)\geq0~,
\end{equation}
where $\mathcal{Z}$ is the set of the two noiseless variables. This limit vanishes if no other variable is coupled to the environment. For example, for $N=3$ with $\varphi_3=0$, only two variables are coupled and $\Omega=0$ along the whole limit. For $q\geq3$, instead, $\Omega$ diverges logarithmically to $+\infty$.
\end{remark}

\begin{remark}[Scope: one shared component plus independent fluctuations]
\label{rem:scope}
The proof uses the dynamics of Eq.~\eqref{Eq. Gen-model} only through the form of the covariance matrix in Eq.~\eqref{Eq. Gen-cov}. This covariance is the sum of two parts. A diagonal matrix $\boldsymbol{D}$ with positive entries describes fluctuations that are independent for each variable, and a term $\boldsymbol{u}\boldsymbol{u}^T$ describes a single shared Gaussian component that enters variable $k$ with weight $u_k$. The conclusion $\Omega\geq0$ therefore holds for any Gaussian vector with covariance $\boldsymbol{C}=\boldsymbol{D}+\boldsymbol{u}\boldsymbol{u}^T$, whatever its origin. Indeed, $\Omega$ is unchanged if each variable is rescaled, and rescaling $z_k$ by $D_{kk}^{-1/2}$ reduces the problem to Eq.~\eqref{Eq. Gen-Omega-N} with $a_k^2=u_k^2/D_{kk}$. The result does not, without further argument, extend to several independent shared environments, whose combined contribution to the covariance matrix cannot in general be written as a single term $\boldsymbol{u}\boldsymbol{u}^T$. In the language of Sec.~\ref{Sec. Common profile}, a single shared component contributes a matrix of rank one, whereas several independent environments generally contribute a matrix of rank greater than one, for which the proof no longer applies.
\end{remark}

\begin{remark}[Common profiles and rank one for white-noise environments]
\label{rem:rank}
For general couplings $f_k(t)$ and $A=W$, the It\^{o} isometry gives the environmental part of Eq.~\eqref{Eq. Gen-cov-full} as the Gram matrix $G_{kl}(t)=\int_0^t f_k(s)f_l(s)\,ds$. By the equality case of the Cauchy--Schwarz inequality, $G(t)$ has rank at most one if and only if all $f_k$ are proportional to a single function on $[0,t]$, i.e., if and only if Eq.~\eqref{Eq. common profile} holds there. Its rank is exactly one provided that at least one $f_k$ is nonzero in $L^2[0,t]$, and it vanishes in the trivial case in which all couplings vanish almost everywhere. This is a statement about rank, not about the sign of $\Omega$. In particular, couplings without a common profile need not produce synergy.
\end{remark}

\subsection{Corollaries}
\label{App. Corollaries}

\begin{corollary}[White-noise environment]
\label{cor:whitenoise}
If $A=W$, then $V_{\mathrm{env}}(t)=\int_0^tg^2(s)\,ds$ and $a_k^2(t)=(\varphi_k^2/\theta_k^2)\,\overline{g^2}(t)$, with $\overline{g^2}(t)=t^{-1}\int_0^tg^2(s)\,ds$. Hence $\Omega(t)\geq0$ for every $N\geq3$. For constant coupling ($g\equiv1$), $a_k=\varphi_k/\theta_k$ and $\Omega$ is independent of $t$.
\end{corollary}

\begin{proof}
The It\^{o} isometry gives $V_{\mathrm{env}}$; insert it into Eq.~\eqref{Eq. Gen-ak} and apply Theorem~\ref{thm:nogo}.
\end{proof}

\begin{corollary}[Ornstein--Uhlenbeck environment]
\label{cor:OU}
Let $A(t)=\int_0^tE(s)\,ds$, with $dE=-\gamma E\,dt+\sigma\,dW(t)$, $\gamma,\sigma>0$, $W$ independent of $\{W_k\}$, and $E(0)$ Gaussian (possibly deterministic) and independent of these Wiener processes. Assume that $g\in L^2[0,t]$. Then $V_{\mathrm{env}}(t)=\int_0^t\!\int_0^tg(s)g(r)\,\mathrm{Cov}(E(s),E(r))\,ds\,dr$ is positive whenever $g$ does not vanish almost everywhere on $[0,t]$, and Theorem~\ref{thm:nogo} applies.
\end{corollary}

\begin{proof}
$\tilde{A}(t)=\int_0^tg(s)E(s)\,ds$ is Gaussian. Its variance is at least its variance conditional on $E(0)$, which equals $\sigma^2\int_0^th(u)^2\,du$ with $h(u)=\int_u^tg(s)e^{-\gamma(s-u)}\,ds$. If $h$ vanishes, then $h'=\gamma h-g$ implies that $g=0$ almost everywhere.
\end{proof}

\begin{corollary}[Weak environmental coupling]
\label{cor:weak}
Let $\varepsilon=\max_k a_k^2$. For fixed $N\geq3$ and $\varepsilon\to0$,
\begin{equation}
  \label{Eq. weak-general}
  \Omega=\frac{1}{\ln 2}\sum_{i<j<k}a_i^2a_j^2a_k^2\,\bigl[1+O(\varepsilon)\bigr].
\end{equation}
\end{corollary}

\begin{proof}
Write $\Omega=\frac{1}{2}\log_2\left[1+(b-c)/c\right]$, with $b$ and $c$ as in the proof of Theorem~\ref{thm:nogo}. Since $S\leq N\varepsilon$, $c=1+O(\varepsilon)$. In Eq.~\eqref{Eq. diff-expanded}, the $j=3$ term is $e_3(1+S)^{N-3}(2+S)=2e_3[1+O(\varepsilon)]$. Each term of $e_j$ with $j\geq4$ is a product of $j$ distinct $a_k^2$, which is at most $\varepsilon^{j-3}$ times a product of three of them, so that $e_j\leq\binom{N-3}{j-3}\varepsilon^{j-3}e_3$ and the terms with $j\geq4$ are $e_3\,O(\varepsilon)$. Hence $(b-c)/c=2e_3[1+O(\varepsilon)]$. Since $e_3=O(\varepsilon^3)$, the expansion $\log_2(1+y)=y[1+O(y)]/\ln2$ gives $\Omega=(e_3/\ln2)[1+O(\varepsilon)]$, which is Eq.~\eqref{Eq. weak-general} because $e_3=\sum_{i<j<k}a_i^2a_j^2a_k^2$. Finally, any three variables considered in isolation have a covariance matrix of the same form, with the same $a_i,a_j,a_k$, so applying Eq.~\eqref{Eq. weak-general} with $N=3$ shows that each term of the sum is, to leading order, the O-information of the corresponding triple.
\end{proof}

\begin{corollary}[Equal coupling ratios]
\label{cor:equal}
If $a_k^2=x\geq0$ for all $k$, then
\begin{equation}
 \label{Eq. equal-couplings-N}
 \Omega(x)=\frac12\log_2\!\left[\frac{(1+Nx)^{N-2}(1+x)^N}{[1+(N-1)x]^N}\right],
\end{equation}
with $\Omega(x)=\binom{N}{3}x^3/\ln2+O(x^4)$ as $x\to0^+$ and
\begin{equation}
 \Omega(x)=\frac{N-2}{2}\log_2 x+\frac12\log_2\!\left[\frac{N^{N-2}}{(N-1)^N}\right]+o(1)
\end{equation}
as $x\to\infty$.
\end{corollary}

\begin{proof}
Setting $S=Nx$ in Eq.~\eqref{Eq. Gen-Omega-N} gives Eq.~\eqref{Eq. equal-couplings-N}. The weak-coupling limit is Corollary~\ref{cor:weak} with $e_3=\binom N3x^3$; the strong-coupling limit follows from the leading power of $x$ in each factor.
\end{proof}

\section{O-information of three Gaussian variables}
\label{App. Gaussian O-info}

This appendix specializes Lemma~\ref{lem:gaussian} to three variables and derives the properties of the O-information in correlation space used in Sec.~\ref{Sec. Toy noninteracting}.

For $N=3$, the diagonal entries $C_{11}$, $C_{22}$, and $C_{33}$ of $\boldsymbol{C}$ are the variances of the three variables, and the off-diagonal entries can be written as $C_{kl}=\rho_{kl}\sqrt{C_{kk}C_{ll}}$ in terms of the correlation coefficients $\rho_{kl}$. Then $\det\boldsymbol{C}=C_{11}C_{22}C_{33}\,(1+2\rho_{12}\rho_{13}\rho_{23}-\rho_{12}^2-\rho_{13}^2-\rho_{23}^2)$ and $\det\boldsymbol{C}_{-k}=C_{ii}C_{jj}(1-\rho_{ij}^2)$, with $\{i,j\}$ the complement of $k$. All variances cancel in Eq.~\eqref{Eq. Gaussian Omega N}, which gives Eq.~\eqref{Eq. Gaussian O-info 5 MT}.

The quantity $g^{\ast}=1+2\rho_{12}\rho_{13}\rho_{23}-\rho_{12}^2-\rho_{13}^2-\rho_{23}^2$ is the determinant of the correlation matrix $\boldsymbol{R}$, with $\boldsymbol{C}=\boldsymbol{D}_C^{1/2}\boldsymbol{R}\,\boldsymbol{D}_C^{1/2}$ and $\boldsymbol{D}_C=\mathrm{diag}(C_{11},C_{22},C_{33})$, so that $\det\boldsymbol{C}=C_{11}C_{22}C_{33}\,g^{\ast}$. Since $\boldsymbol{C}$ and $\boldsymbol{R}$ are related by a congruence, $\boldsymbol{C}$ is positive definite if and only if $\boldsymbol{R}$ is. By Sylvester's criterion~\cite[Theorem~7.2.5]{HornJohnson2013}, this requires the leading principal minors of $\boldsymbol{R}$, namely $1$, $1-\rho_{12}^2$, and $g^{\ast}$, to be positive. For $(\rho_{12},\rho_{13},\rho_{23})\in[-1,1]^3$, the condition $g^{\ast}>0$ already implies $|\rho_{12}|<1$, because $|\rho_{12}|=1$ gives $g^{\ast}=-(\rho_{13}-\rho_{12}\rho_{23})^2\leq0$. Hence $\boldsymbol{C}$ is positive definite if and only if $g^{\ast}>0$.

Eq.~\eqref{Eq. Gaussian O-info 5 MT}, as well as $g$ and $g^{\ast}$, are invariant under any permutation of the three correlation coefficients and under sign changes of any two of them, $(\rho_{12},\rho_{13},\rho_{23})\to(\epsilon_1\rho_{12},\epsilon_2\rho_{13},\epsilon_3\rho_{23})$ with $\epsilon_i=\pm1$ and $\epsilon_1\epsilon_2\epsilon_3=1$, since they depend on the correlations only through their squares and the product $\rho_{12}\rho_{13}\rho_{23}$. Physically, permutations correspond to relabeling the variables, and a sign change of two correlations corresponds to reversing the sign of the variable they share, $z_k\to-z_k$. Neither operation changes $\Omega$. These symmetries explain the symmetric structure of the slices in Fig.~\ref{Fig_0.1}(a,b).

\section{Exact covariance for linear interactions}
\label{App. Linear}

This appendix derives the exact covariance matrix, and hence the O-information, of the toy model of Sec.~\ref{Sec. Toy model} when only the linear interaction $m_1$ is present. We first recall the standard solution of a general linear system with additive noise (App.~\ref{App. Linear general}), then specialize it to the toy model (App.~\ref{App. Linear toy}).

\subsection{General linear systems with additive noise}
\label{App. Linear general}

For linear interactions, $F_k(\boldsymbol{z})=\sum_l B_{kl}z_l$, where the constant $B_{kl}$ measures how $z_l$ feeds back on the drift of $z_k$. With a white-noise environment, Eq.~\eqref{Eq. Gen-model} then reads
\begin{equation}
  \label{Eq. Linear SDE}
  d\boldsymbol{z} = (\boldsymbol{\mu}+\boldsymbol{B}\boldsymbol{z})\,dt + \boldsymbol{\Theta}\,d\boldsymbol{W}_{\mathrm{loc}}(t) + \boldsymbol{f}(t)\,dW(t)~,
\end{equation}
with $\boldsymbol{\Theta}=\mathrm{diag}(\theta_1,\ldots,\theta_N)$ and $\boldsymbol{W}_{\mathrm{loc}}=(W_1,\ldots,W_N)^T$. This is a multivariate Ornstein--Uhlenbeck process with time-dependent noise amplitudes~\cite[Ch.~6]{Sarkka2019}.

\textit{Solution.} For a single variable, $dx=bx\,dt+\sigma\,dW$, multiplying by the integrating factor $e^{-bt}$ gives $d(e^{-bt}x)=e^{-bt}\sigma\,dW$, hence $x(t)=e^{bt}x(0)+\int_0^te^{b(t-s)}\sigma\,dW(s)$: each noise kick received at time $s$ is propagated to time $t$ by the deterministic dynamics. The same steps with the matrix exponential $e^{\boldsymbol{B}t}$ give~\cite[Secs.~3.3 and 4.3]{Sarkka2019}
\begin{align}
  \boldsymbol{z}(t) = {} & e^{\boldsymbol{B}t}\boldsymbol{z}(0) + \int_0^t e^{\boldsymbol{B}(t-s)}\boldsymbol{\mu}\,ds \notag\\
  & + \int_0^t e^{\boldsymbol{B}(t-s)}\left[\boldsymbol{\Theta}\,d\boldsymbol{W}_{\mathrm{loc}}(s)+\boldsymbol{f}(s)\,dW(s)\right].
  \label{Eq. Linear solution}
\end{align}
The first line is deterministic and gives the mean, which does not affect $\Omega$ (Lemma~\ref{lem:gaussian}). The second line contains all fluctuations.

\textit{Gaussianity.} The stochastic integrals in Eq.~\eqref{Eq. Linear solution} have deterministic integrands. Each is the limit of sums $\sum_i g(s_i)\,[W(s_{i+1})-W(s_i)]$ of independent Gaussian increments weighted by fixed numbers. Linear combinations of jointly Gaussian variables are Gaussian, and so are their mean-square limits~\cite{Grimmett2001,Sarkka2019}. Hence $\boldsymbol{z}(t)$ is jointly Gaussian at every time. This argument fails for nonlinear interactions, for which the integrands depend on $\boldsymbol{z}$ itself.

\textit{Covariance.} For deterministic functions $g$ and $h$ and a standard Wiener process $W$, the It\^{o} isometry states that~\cite[Sec.~4.1]{Sarkka2019}
\begin{equation}
  \label{Eq. Ito isometry}
  \mathbb{E}\!\left[\int_0^t g(s)\,dW(s)\int_0^t h(s)\,dW(s)\right]=\int_0^t g(s)h(s)\,ds~.
\end{equation}
Intuitively, the increments $dW(s)=W(s+ds)-W(s)$ over non-overlapping intervals are uncorrelated, and each has variance $ds$. Only products $g(s)h(s)$ at equal times therefore contribute to the expectation, each with weight $ds$. Integrals over independent Wiener processes are uncorrelated. Applying Eq.~\eqref{Eq. Ito isometry} to each noise source in Eq.~\eqref{Eq. Linear solution} and writing $\tau=t-s$ gives
\begin{equation}
  \label{Eq. Linear cov}
  \boldsymbol{C}(t) = \int_0^t e^{\boldsymbol{B}\tau}\left[\boldsymbol{\Theta}^2+\boldsymbol{f}(t-\tau)\boldsymbol{f}^T(t-\tau)\right]e^{\boldsymbol{B}^T\tau}\,d\tau~.
\end{equation}
Equivalently, differentiating with respect to $t$, $\boldsymbol{C}$ solves $d\boldsymbol{C}/dt=\boldsymbol{B}\boldsymbol{C}+\boldsymbol{C}\boldsymbol{B}^T+\boldsymbol{\Theta}^2+\boldsymbol{f}\boldsymbol{f}^T$ with $\boldsymbol{C}(0)=0$~\cite[Sec.~3.3 and Ch.~6]{Sarkka2019}. Without interactions ($\boldsymbol{B}=0$), Eq.~\eqref{Eq. Linear cov} reduces to Eq.~\eqref{Eq. Fokker-Planck 2 MT}.

\subsection{Toy model with a linear interaction}
\label{App. Linear toy}

For the toy model with $m_2=m_3=0$, only $z_1$ receives the feedback $m_1(z_1+z_2+z_3)$. Hence $\boldsymbol{B}=m_1\,\boldsymbol{e}_1\boldsymbol{1}^T$, the matrix whose first row is $(m_1,m_1,m_1)$ and whose other rows vanish, with $\boldsymbol{e}_1=(1,0,0)^T$ and $\boldsymbol{1}=(1,1,1)^T$.

\textit{Matrix exponential.} Since $\boldsymbol{1}^T\boldsymbol{e}_1=1$, we have $\boldsymbol{B}^2=m_1\boldsymbol{B}$ and $\boldsymbol{B}^n=m_1^{n-1}\boldsymbol{B}$ for $n\geq1$. Summing the exponential series,
\begin{equation}
  e^{\boldsymbol{B}\tau} = \mathbb{I} + \sum_{n\geq1}\frac{m_1^{n-1}\tau^n}{n!}\,\boldsymbol{B} = \mathbb{I} + \frac{\chi(\tau)}{m_1}\,\boldsymbol{B}~,
\end{equation}
with $\chi(\tau)=e^{m_1\tau}-1$.
Acting on a vector $\boldsymbol{v}$, this gives $e^{\boldsymbol{B}\tau}\boldsymbol{v}=\boldsymbol{v}+\chi(\tau)\,(v_1+v_2+v_3)\,\boldsymbol{e}_1$. In words, a noise kick $\boldsymbol{v}$ leaves $z_2$ and $z_3$ unchanged, since they receive no feedback, whereas its effect on $z_1$ is modified in proportion to the total kick $v_1+v_2+v_3$.

\textit{Propagated noise vectors.} The local noise of $z_k$ enters along $\theta_k\boldsymbol{e}_k$, and the shared noise along $\boldsymbol{\varphi}$ (constant coupling). After a time $\tau$, these vectors become
\begin{align*}
  \theta_1\boldsymbol{e}_1 &\to \theta_1\,e^{m_1\tau}\,(1,0,0)^T~,\\
  \theta_2\boldsymbol{e}_2 &\to \theta_2\,\bigl(\chi(\tau),1,0\bigr)^T~,\\
  \theta_3\boldsymbol{e}_3 &\to \theta_3\,\bigl(\chi(\tau),0,1\bigr)^T~,\\
  \boldsymbol{\varphi} &\to \bigl(\varphi_1+\chi(\tau)\,\Phi,\varphi_2,\varphi_3\bigr)^T~,
\end{align*}
with $\Phi=\varphi_1+\varphi_2+\varphi_3$. By Eq.~\eqref{Eq. Linear cov}, $\boldsymbol{C}(t)$ is the time integral of the sum of the outer products of these four vectors. Only three elementary integrals appear:
\begin{align*}
  I_1(t) &= \int_0^t \chi(\tau)\,d\tau = \frac{e^{m_1t}-1}{m_1}-t~,\\
  J(t) &= \int_0^t e^{2m_1\tau}\,d\tau = \frac{e^{2m_1t}-1}{2m_1}~,\\
  I_2(t) &= \int_0^t \chi^2(\tau)\,d\tau = J(t)-2\,\frac{e^{m_1t}-1}{m_1}+t~.
\end{align*}

\textit{Result.} The covariances involving $z_1$ are
\begin{align}
  C_{11} &= \theta_1^2\,J + \left(\theta_2^2+\theta_3^2+\Phi^2\right) I_2 + 2\varphi_1\Phi\, I_1 + \varphi_1^2\, t~, \notag\\
  C_{1l} &= \left(\theta_l^2+\varphi_l\Phi\right) I_1 + \varphi_1\varphi_l\, t~, \qquad l=2,3~,
  \label{Eq. Linear cov toy}
\end{align}
while $C_{22}$, $C_{33}$, and $C_{23}$ are unaffected by the interaction and are given by Eq.~\eqref{Eq. Gen-cov} with $V_{\mathrm{env}}=t$. As a check, for $m_1\to0$ one has $I_1,I_2\to0$ and $J\to t$, which recovers the non-interacting result. The O-information follows by inserting these covariances into Eq.~\eqref{Eq. Gaussian O-info 5 MT}; the exact curves in Fig.~\ref{Fig_0.2} were obtained in this way.

\textit{Why Theorem~\ref{thm:nogo} no longer applies.} Without interactions, the only off-diagonal covariances are $\varphi_k\varphi_l\,t$, which form a rank-one matrix. For $m_1\neq0$, the terms $\theta_l^2 I_1$ in $C_{1l}$ show that the local noises of $z_2$ and $z_3$, transmitted to $z_1$ through the feedback, now contribute to their correlations with $z_1$. In addition, the shared signal enters $z_1$ with the time-dependent weight $\varphi_1+\chi(\tau)\Phi$, which plays the role of a coupling profile that differs from those of $z_2$ and $z_3$. Both effects destroy the structure on which Theorem~\ref{thm:nogo} relies, namely independent fluctuations plus a single shared component, which is why synergy becomes possible.

\providecommand{\noopsort}[1]{}\providecommand{\singleletter}[1]{#1}
%


\end{document}